\documentclass[onecolumn,amsmath,amssymb,aps]{revtex4-2}
\usepackage{color}
\usepackage{graphicx}
\usepackage{amsmath}
\usepackage{dcolumn}
\usepackage{bm}
\usepackage{subfigure}
\usepackage[english]{babel}
\usepackage{amscd}
\usepackage{epsfig}
\usepackage{tabularx}
\usepackage{graphicx}
\usepackage{latexsym}
\usepackage{amsmath}
\usepackage{amsfonts}
\usepackage{amssymb}
\usepackage[latin1]{inputenc}
\usepackage{times}
\usepackage[T1]{fontenc}
\usepackage{latexsym}
\usepackage{graphics}
\usepackage{verbatim}
\usepackage[absolute]{textpos}
\usepackage{wrapfig,times}
\usepackage{amsthm}
\usepackage{setspace}
\usepackage{color}
\definecolor{Red}{rgb}{0.9,0.1,0.1}
\definecolor{blue}{rgb}{0.25,0.25,0.6}

\newcommand{\be}{\begin{equation}}
\newcommand{\ee}{ \end{equation}}
\newcommand{\ben}{\begin{eqnarray}}
\newcommand{\een}{\end{eqnarray}}

\begin{document}

\title{Intermediate statistics: addressing the thermoelectric properties of solids}

\author{André A. Marinho$^{1,2}$, Francisco A. Brito$^{2,4}$, G.M. Viswanathan$^{1,3}$, C.G. Bezerra$^{1}$}

\affiliation{$^{1}$ Departamento de Física, Universidade Federal do Rio
Grande do Norte, 59078-900 Natal, RN, Brazil
\\
$^{2}$ Departamento de Física, Universidade Federal de
Campina Grande, 58109-970 Campina Grande, Paraíba, Brazil
\\
$^3$National Institute of Science and Technology
                   of Complex Systems,
                   Universidade Federal do Rio Grande do Norte,
                   Natal--RN, 59078-900, Brazil
\\
$^{4}$ Departamento de Física, Universidade Federal da Paraíba,
Caixa Postal 5008, 58051-970 João Pessoa, Paraíba, Brazil}
\date{\today}

\begin{abstract}

We study the
    thermodynamics of a crystalline solid by applying intermediate
    statistics obtained by deforming known solid state models using
    the mathematics of $q$-analogs.  We apply the resulting
    $q$-deformation to both the Einstein and Debye models and study
    the deformed thermal and electrical conductivities and the
    deformed Debye specific heat. We find that the $q$-deformation
    acts in two different ways --- but not necessarily as independent
    mechanisms. First, it acts as an effective factor of disorder or
    impurity, modifying the characteristics of a crystalline
    structure, which are phenomena described by q-bosons. Second, it
    also manifests intermediate statistics, namely, the B-anyons (or
    B-type systems). For the latter case, we have identified the
    Schottky effect, normally associated with high-$T_c$
    superconductors in the presence of rare-earth-ion impurities.  We
    also find that it increases the specific heat of the solids beyond the
    Dulong-Petit limit at high temperature. Such an effect is usually
    related to anharmonicity of interatomic
    interactions. Alternatively, since in the $q$-bosons case the
    statistics are in principle maintained, the effect of the
    deformation acts more slowly due to a small change in the crystal
    lattice. On the other hand, B-anyons that belong to modified
    statistics are more sensitive to the deformation. The
results reported here may be verified experimentally, for instance, in experimental samples by inserting
impurities, or changes in pressure or temperature if one assumes these tuning quantities are related
with the q-deformation parameter.

\end{abstract}


\maketitle

\section{Introduction}
\label{int}


Studies conducted in the second half of the 20th century have
demonstrated that the presence of defects or impurities in a crystal
perturbs the local electrostatic potential, thereby breaking the
translational symmetry of the periodic potential \cite{and,ell,lee}. This disturbance can produce electronic wave
functions located near the impurity failing to propagate through the
crystal. The process of inserting impurities made out of known
elements in a semiconductor (i.e., doping) is a well known method for
obtaining significant changes in transport properties of crystalline
solids.

On the other hand, extensive research on $q$-deformation via quantum groups and quantum
algebras has been undertaken across various fields of physics,
including cosmology and condensed matter
\cite{lei,che,ien,ler,fuc,wil,bie,mac,lee2,sin,rpv,chai,col,sou,lav5,bri5,Marinho:2019zny,Marinho:2019zny2,
  chu,chu1,chu2,hou,pra,bou}. Even in the absence of a universally
recognized satisfactory definition of a quantum group, all the present
proposals suggest the idea of deforming a classical object which may
be, for example, an algebraic group or a Lie group. In
  such proposals, the deformed objects may lose the group property
  \cite{arik,jim}.
However, this is not always the case, for instance we cite one-parameter and two-parameter deformed fermion and boson quantum groups
  \cite{ubr,aba2,nn1,nn2,nn3}.

The literature also includes studies on intermediate or fractional
statistics, which describe anyons
\cite{aro,sen,ach,fiv,per,dal,nar,lav2,ams,lav,chungnew1,lav3,aba5,man,rov,free,aba,sid,cla,shen,gli,lens,Morier,greiter,kwan}. The
latter are one of the versions of nonstandard quantum statistics \cite{gen,gre,pol,lav1,kha}. The results
provides that $q$-deformation, associated to deformed quantum statistics, acts in two different ways but not necessarily as independent mechanisms: it acts as a factor of disorder or impurity, modifying the characteristics of a crystalline structure, which are phenomena described by either the $q$-bosons or the quantum group invariant ones, and also as a manifestation of intermediate statistics, the $B$-anyons (or $B$-type systems). The statistical mechanics of anyons have been studied in two spatial dimensions using a natural distinction existing between boson-like and fermion-like anyons \cite{wil}. Anyons also manifest in any dimensions, including ``spinon'' excitations in one-dimensional antiferromagnets, as shown long ago by Haldane \cite{haldane}. Fractional statistics determined by the relation defining permutation symmetry in the many-body wave function
\be\label{eq}\psi(x_1,\ldots x_j\cdots x_i \ldots,
x_n)\!=e^{i\pi\alpha}\psi(x_1,\ldots x_i\ldots x_j \ldots, x_n)~,\ee

\noindent where the statistics determining parameter $\alpha$ is a real number and conventionally described as $0\leq\alpha\leq 1$ under the case of fractional statistics.
The limits $\alpha=0, 1$ correspond to bosons and fermions, respectively. From the literature, we know that the permutation symmetry is related to rotations
in two spatial dimensions.

In this paper, we use the mathematics of $q$-analogs to study a family of deformed models. Specifically, we apply the $q$-deformation in solids in a way that has similar effects to those of defects or impurities in the Einstein and Debye model of solids \cite{bri,bri1,bri2,bri3}. Our results show that the $q$-deformation parameter modifies the thermodynamic quantities such as entropy, specific heat, as well as thermal and electric conductivities \cite{zim}. However, at this point we should answer a timely question that naturally arises: why is the $q$-deformation a good approach? The answer is simple: the theory of $q$-series has centuries of solid mathematical underpinning.  The study of $q$-series began with some theorems of Gauss (e.g., for theta functions), Euler (e.g., the combinatorial version of the pentagonal number theorem) and Cauchy (e.g.,
$q$-analog of the binomial theorem). At the turn of the 19th century and the begininng of the 20th century, the English mathematicians F.~H.~Jackson (known for the Jackson derivative) and L.~J.~Rogers (known for the Rogers-Ramanujan identities) built the foundations of this important research area of mathematics. Given this long mathematical tradition, the $q$-analogs have a far more solid theoretical basis in comparison to other deformations of the standard algebra.

The paper is organized as follows. In Sec.\ \ref{ais} we introduce the $q$-deformed algebra and develop the intermediate statistics. For the convenience of the reader, and to help us to make comparisons with the $B$-anyons statistics, we revisit the $q$-deformed boson-like statistics \cite{bri}. In Sec.\ \ref{iis} we apply the the $B$-anyons statistics in the Einstein and Debye solid models in three dimensions.  Finally, in Sec.\ \ref{con} we summarize the main results obtained in this work and make our final comments.

\section{The Algebra of Interpolating Statistics}
\label{ais}
To investigate boson-like interpolating statistics we will examine the $q$-deformed algebra studied by Biedenharn and Macfarlane \cite{bie,mac}. This generalization inserts a parameter $q$, that can be a real or complex
number, defined by
\be \label{e.1} bb^{\dagger} - qb^{\dagger}b = q^{-N},\qquad 0\leq q \leq 1, \ee
in terms of creation and annihilation operators $b^{\dagger}$ and $b$, respectively, and quantum number $N$.
The regime $q = 1$ corresponds to the Bose-Einstein limit. We also have the following relations
\begin{align}
  [N,b]&=-b~, \nonumber \\
  [N,b^{\dagger}]& =b^{\dagger}~, \nonumber \\
  [N]& =b^{\dagger}b~, \nonumber\\
  [N+1]& =q[N]+q^{-N}~,
  \label{eq0}
\end{align}

\noindent where, following a long tradition, the basic number $[N]$ 
is defined as 
\cite{erns}
\begin{equation}\label{eq2} [N]=\frac {q^{{N}}-q^{-{N}}}{q-q^{-{1}}}.\end{equation}

\noindent On the other hand, one may transform the $q$-Fock space into the configuration space (Bargmann holomorphic
representation) \cite{flo} as
\begin{eqnarray} \label{e32}b^{\dagger} \Rightarrow x,\qquad\qquad b \Rightarrow \partial_{x}^{q},\end{eqnarray}
where $\partial_{x}^{q}$ is the Jackson derivative (JD) \cite{jac} given by
\begin{equation}\partial_x^{q}f{(x)}=\frac {f{(qx)}- f{(q^{-1}x)}}{x{(q-q^{-1})}}.\end{equation}
Note that JD becomes an ordinary derivative as $q\to1$. Therefore, JD naturally occurs in quantum deformed structures and it
turns out to play a crucial role in $q$-generalization of thermodynamic relations \cite{lav1}.
The $q$-Fock space spanned by the orthonormalized eigenstates $|n \rangle$ is constructed according to
\begin{eqnarray} {|n\rangle} =\frac{(b^{\dagger})^{n}} {\sqrt{[n]!}}{|0\rangle},\qquad\qquad b{|0\rangle}=0,\end {eqnarray}
where the basic factorial is given by $[n]!=[n][n-1][n-2]\cdots 1$, with $n$ corresponding to
any non-negative integer.
The actions of the operators $b$, $b^{\dagger}$ and $N$ on the states $|n\rangle$ in the $q$-Fock space are well-known and read
\begin{equation} b^{\dagger}{|n\rangle} = (n+1)^{1/2} {|n+1\rangle},\end{equation}
\begin{equation} b{|n\rangle} = (n)^{1/2} {|n-1\rangle},\end{equation}
and
\begin{equation} N{|n\rangle} = n{|n\rangle}.\end{equation}

\subsection{$q$-bosons}

For the convenience of the reader, and to help us to make comparisons, let us revisit the $q$-deformed occupation number $n_{i}^{q}$ for a boson-like statistics \cite{bri}. We start with the Hamiltonian of quantum harmonic oscillators
\cite{bie, mac, nar,ach,lav2,ams,lav,chungnew1,wil}

\be\label{eq1}{\cal H} = \sum_{i}{(E_i-\mu)}{N_i},\ee

\noindent where $\mu$ is the chemical potential. 
The $q$-deformed Hamiltonian depends implicitly on $q$ through the number operator defined in Eqs.\ (\ref{eq0}) and (\ref{eq2}).

We know from standard statistical mechanics that the expectation value of the occupation number is given by
\begin{align} &  [n_i]= \nonumber \\
 &   \frac{\exp\left[-\beta(E_i-\mu)\right]{\rm tr}\left\{\exp\left[-\beta\displaystyle\sum_{n_j}
{(E_j-\mu)n_j b_{i}b_{i}^\dagger }\right]\right\}}{{\rm tr}\left\{\exp\left[-\beta\displaystyle\sum_{n_j}{(E_j-\mu)n_j }
\right]\right\}}. \end{align}
We consider the definition $b_i b_{i}^\dagger=[1+n_i]$ and use the cyclic property of the trace to find \cite{lee2}
\begin{align} \label{e37}[n_i] &= \exp\left[-\beta(E_i-\mu)\right][1+n_i]\,\, \rightarrow \,\, \frac{[n_i]}
{[1+n_i]} = z e^{-\beta E_i}, \nonumber \\  \end{align}
where $z=\exp(\beta\mu)$ is the fugacity. Proceeding further, we obtain the equation for the $q$-deformed
mean occupation number as \cite{lav2},
\begin {equation}\label{e38} n_{i}^{q} = \frac{1}{2\ln q}\ln\left[\frac{z^{-1}\exp(\beta E_i)-q^{-1}}{z^{-1}
    \exp(\beta E_i)-q}\right].\end{equation}
The explicit steps to obtain Eq.\ (\ref{e38}) are given in appendix A.

Interestingly, as an alternative approach, the $q$-deformed algebra
allows us to derive Eq.\ (\ref{e38}) using the JD. This is
possible because we can formally express the total number of
$q$-deformed particles in the form
\be\label{eqXi} N^{q}=zD_{z}^{q}\ln{\Xi}\equiv \displaystyle\sum_{i}n_{i}^{q},\ee
where $D_{z}^{q}$ is a $q$-deformed differential operator defined as
\ben D_{z}^{q}=\frac{q-q^{-1}}{2\ln{q}}
\partial_{z}^{q} \quad\mbox{(when\:\: $q\to 1$,}\quad D^q_{z} \to \frac{\partial}{\partial z}),\een
and $\ln\Xi$ defines the undeformed grand-partition function.

\subsection{B-anyons}
\label{Banyons}

Let us now derive a distribution function for B-anyons (or B-type
systems). We remark that from now on we refer $\bar{q}$-deformation to $B$-anyons and $q$-deformation to $q$-bosons along  the remainder of this paper. For B-anyons systems, we will expand Eq.\ (\ref{e38}) in the
following power series \cite{ams,lav,lav3} 
\begin{equation}\label{ex} n_{i}^{\bar{q}} = \frac{1}{y}+\left(\frac{1}{6y}+\frac{\bar{q}}{2y^2}+\frac{1}{3y^3}\right)\varepsilon^2+
\cdots,\end{equation}
where $y=\exp(\eta)-\bar{q}$, $\eta=\beta(E_i-\mu)$ and $\bar{q}=1-\varepsilon$, with $\varepsilon\ll 1$. 
In Refs.\ \cite{ams,lav,lav3} we find the expansion of Eq.\ (\ref{ex}) to higher orders. High order terms of expansion of Eq.\ (\ref{ex}) can be associated to interactions between particles, such as van der Waals, three-body cluster, and quantum scattering \cite{ams}. In the present work, once our goal is to make a comparison between $q$-bosons and $B$-anyons, we take only the main term, i.e.
\begin {equation}\label{e50} n_{i}^{\bar{q}}=\frac{1}{y}=\frac{1}{z^{-1}\exp(\beta E_i)-\bar{q}}.\end{equation}
In the literature we find important works \cite{ach,lav,aba5} emphasizing that both Eqs.\ (\ref{ex}) and (\ref{e50}) provide several
applications for B-type systems.

Now by simple inspection of Eq.~(\ref{eqXi}), but using ordinary derivative instead of JD, we can write the logarithm of $\bar{q}$-deformed grand-partition function $\Xi_{\bar{q}}$ as
\begin{equation}\label{e50.1}\ln{\Xi_{\bar{q}}}=-\displaystyle\sum_{i}{{\bar{q}\,}^{-1}\ln{\left[1-\bar{q}z\exp(-\beta E_i)\right]}}.
\end{equation}
In the next section, we will deepen our study of solid models by introducing the results obtained in the present section to the energy spectrum in three dimensions, as we can also see in Refs.\ \cite{rov,free}. We must remark that the $q$-bosonic and $\bar{q}$-anyonic behaviors are consequences of the deformed occupation numbers ($n_{i}^{{q}}$ and $n_{i}^{\bar{q}}$), which have a cascading effect on all quantities that depend on them, as we will see later on this work.

\section{Implementation of Interpolating Statistics}
\label{iis}

\subsection {Deformed Einstein solid}
\label{des}
\begin{figure*}[htb]
\centerline{
  \includegraphics[{angle=90,height=10.0cm,angle=270,width=6.1cm}]{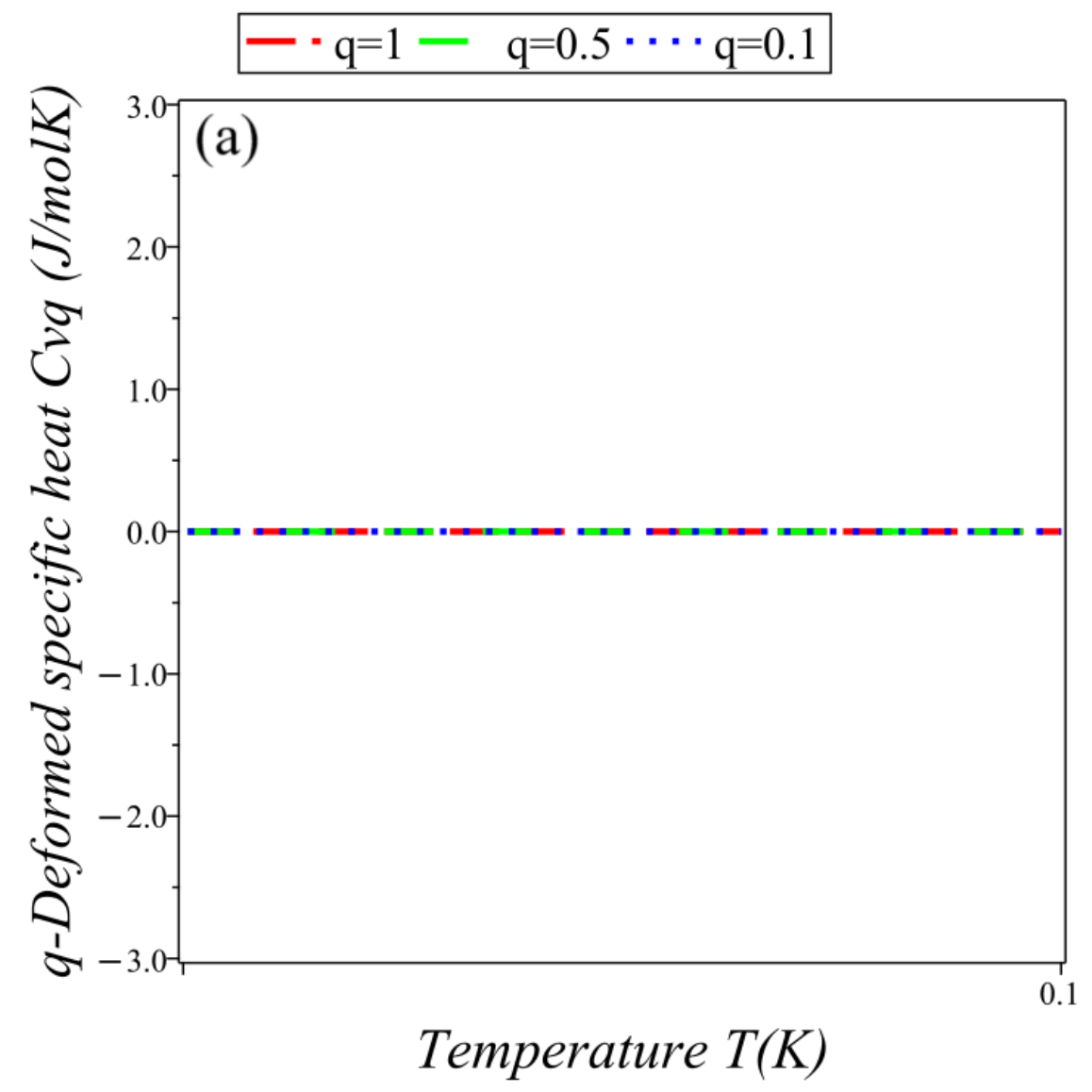}
  \includegraphics[{angle=90,height=10.0cm,angle=270,width=6.1cm}]{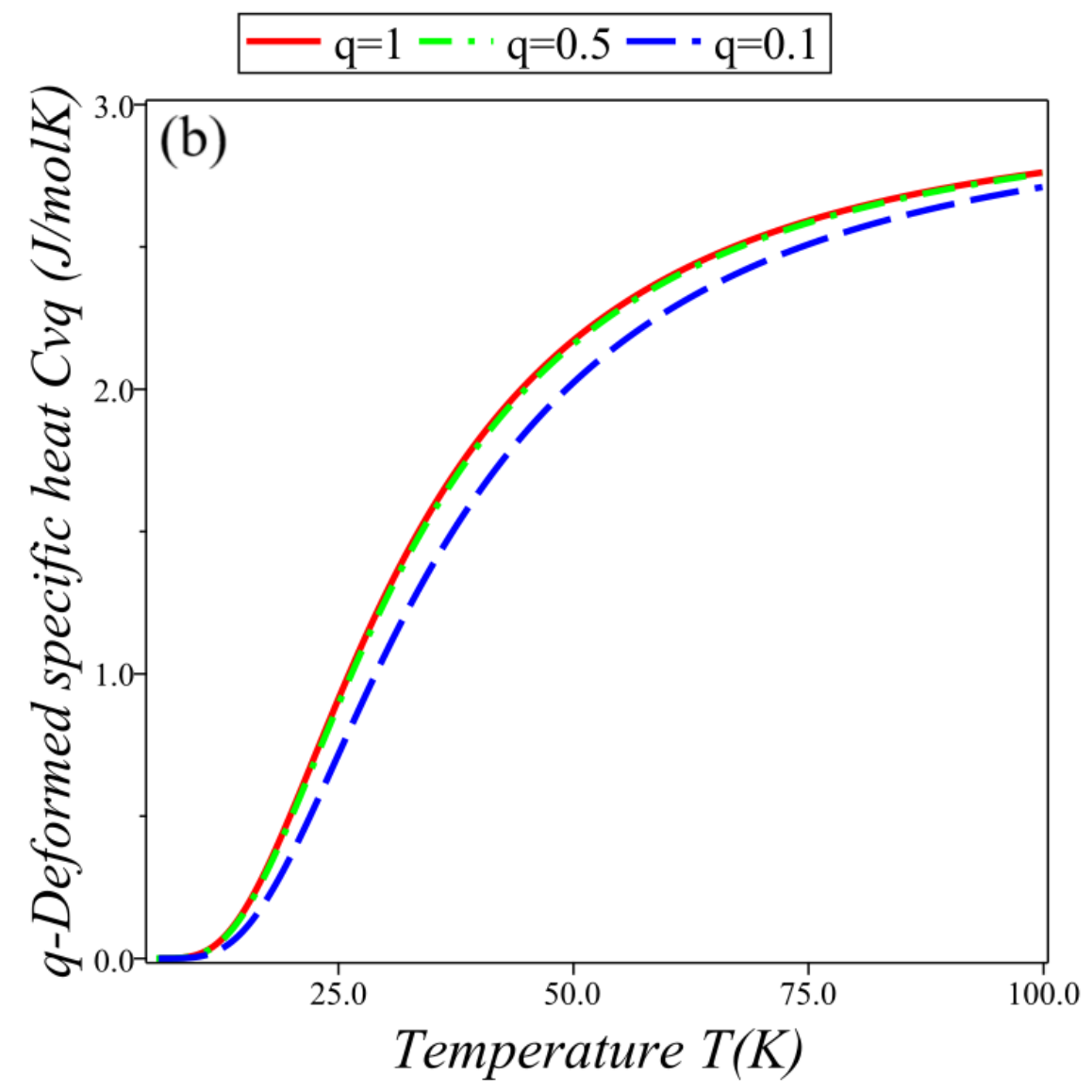}
  \includegraphics[{angle=90,height=10.0cm,angle=270,width=6.1cm}]{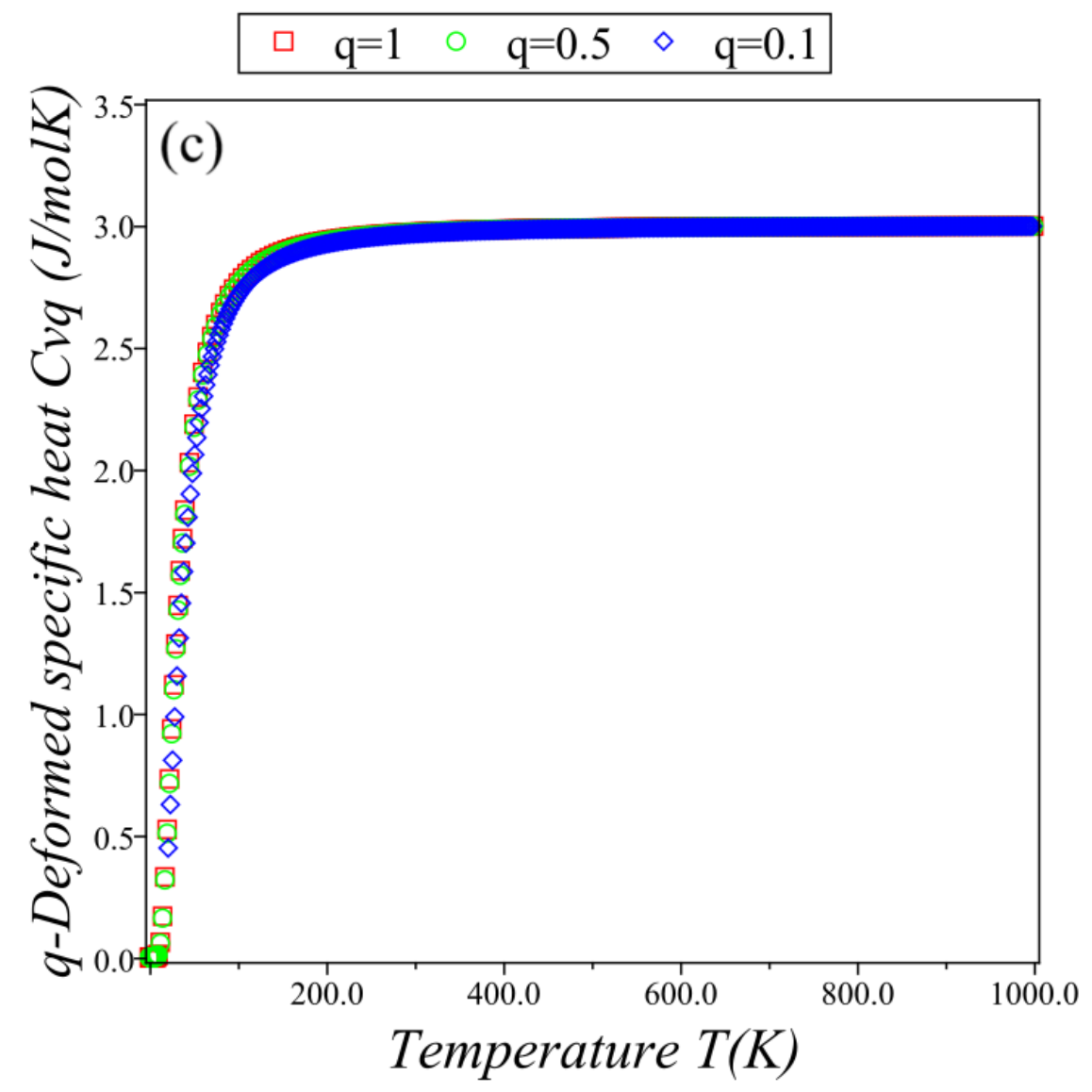}
}\caption{$q$-deformed specific heat $c_{V_{q}}$ vs
    temperature $T$ in the following intervals: (a) $T=0-0.1 K$; (b) $T=0-100 K$; and (c) $T=0-1000 K$. See Ref.\ \cite{bri}. Discussion is in the main text.}\label{grafi2}
\end{figure*}

\begin{figure*}[htb]
\centerline{
\includegraphics[{angle=90,height=10.0cm,angle=270,width=6.1cm}]{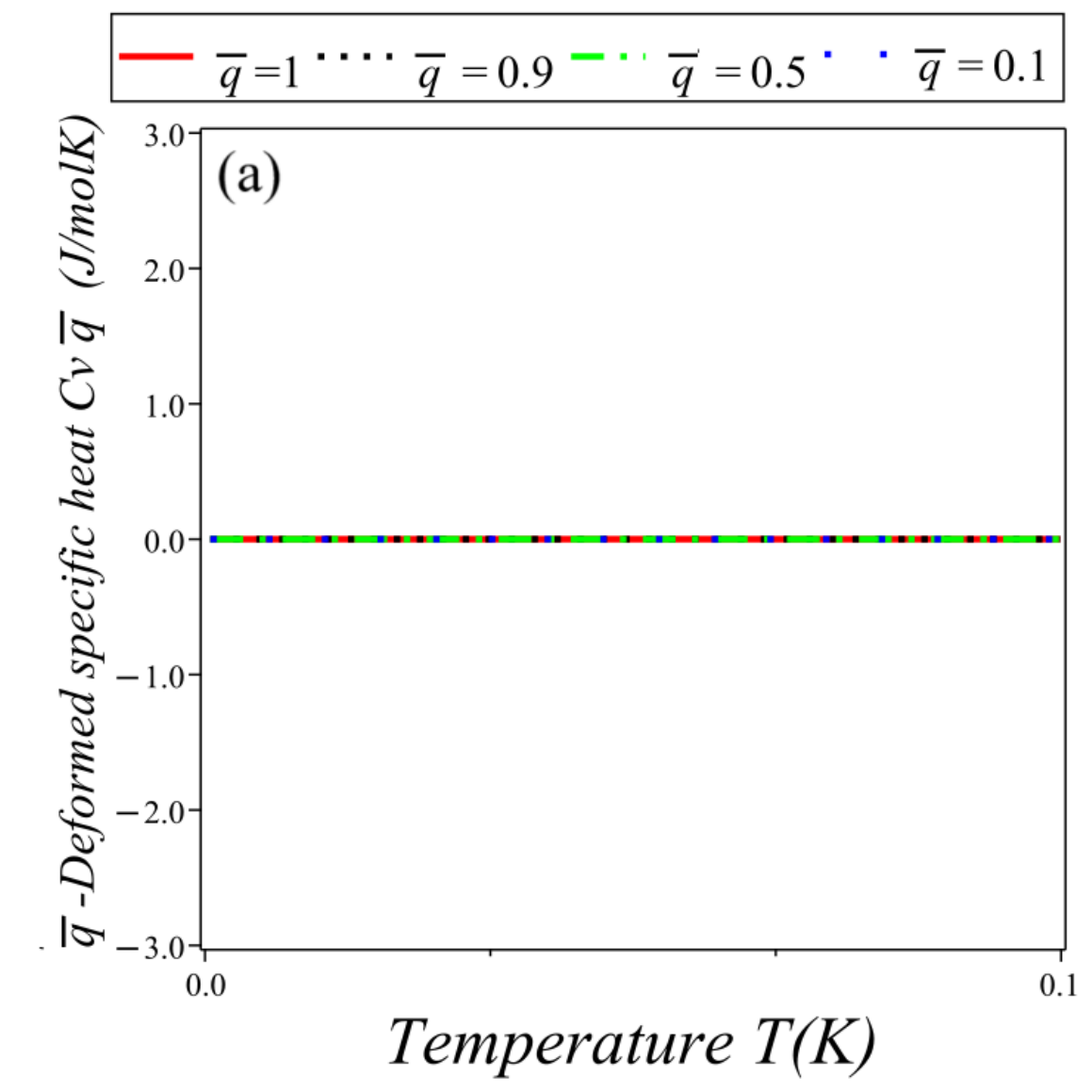}
\includegraphics[{angle=90,height=10.0cm,angle=270,width=6.1cm}]{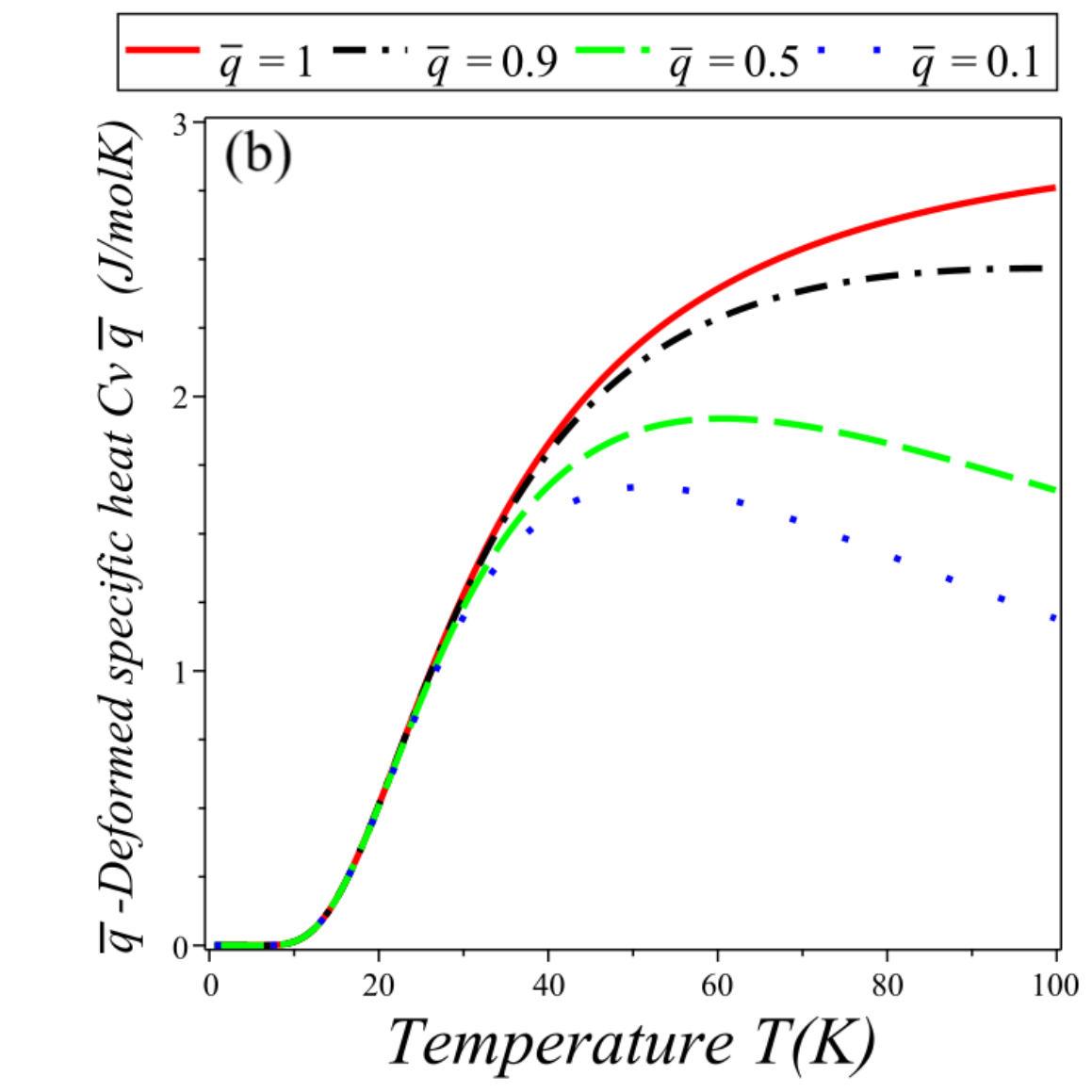}
\includegraphics[{angle=90,height=10.0cm,angle=270,width=6.1cm}]{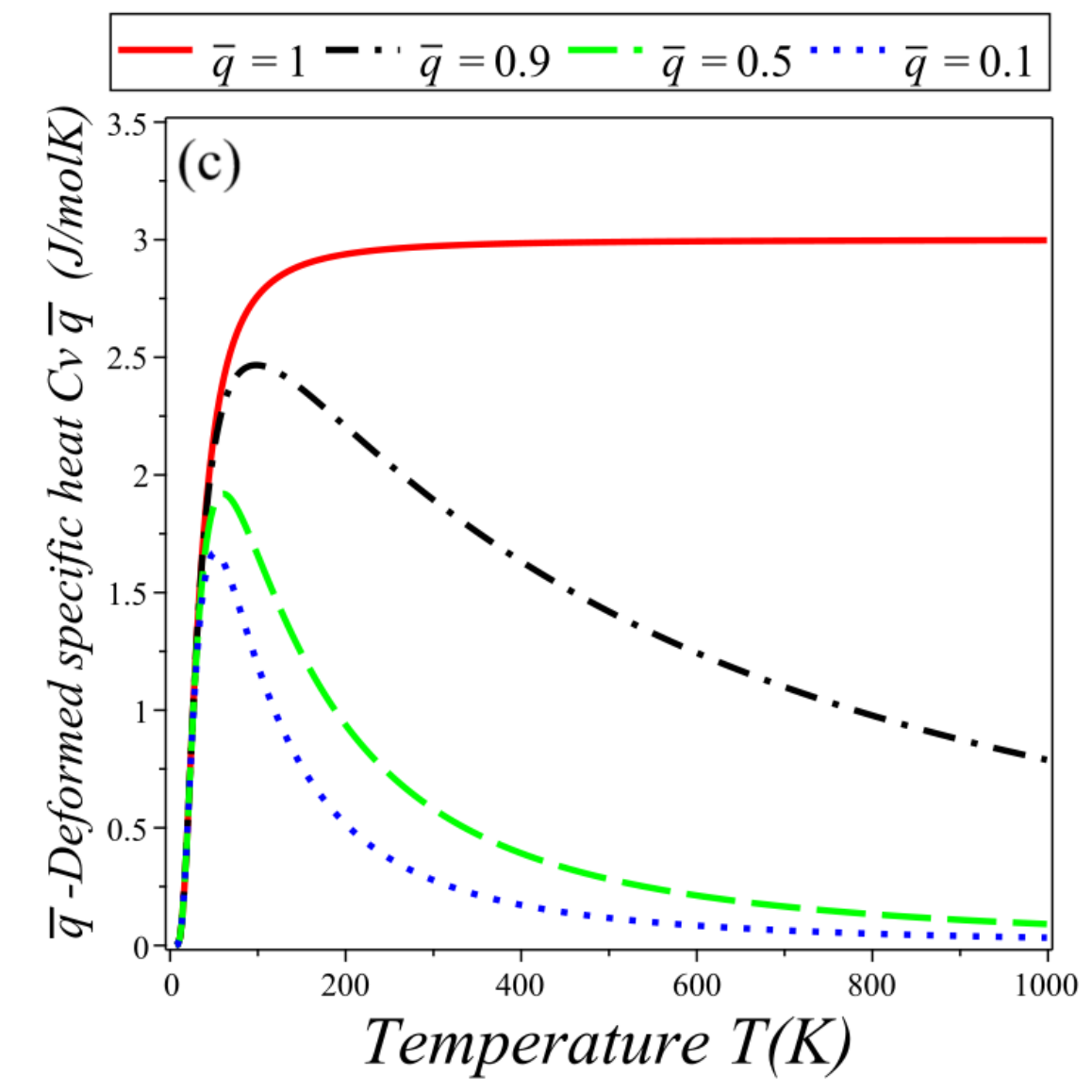}
}\caption{$\bar{q}$-deformed specific heat $c_{V_{\bar{q}}}$ vs
    temperature $T$ in the following intervals: (a) $T=0-0.1 K$; (b) $T=0-100 K$; and (c) $T=0-1000 K$. Discussion is in the main text.}\label{graf2}
\end{figure*}

We can consider the solid in contact with a thermal reservoir at temperature $T$, where we have $n_j$
labeling the $j$-th oscillator. Given a microscopic
state $\{n_j\}=\{n_1,n_2,\ldots,n_N\}$, the energy of this state can be written as
\begin{equation}\label{eq8} E\{n_j\} =
\displaystyle\sum_{j=1}^{3N}{\hbar\omega_E\left(n_j+\frac{1}{2}\right)},\end{equation}
where $\omega_E$ is the Einstein characteristic frequency. 
Now, substituting Eq.\ (\ref{eq8}) into Eq.\ (\ref{e50.1}), excluding the zero point term ${\hbar\omega}/{2}$, we can determine a $\bar{q}$-deformed Helmholtz free energy per
oscillator given by
\begin{align}\label{eq10}
f_{\bar{q}} &= -\frac{1}{\beta}\displaystyle\lim_{N\to\infty}\frac{1}{N}\ln\Xi_{\bar{q}}
=\frac{k_{B} T}{\bar{q}}\ln(1-\alpha_{\bar{q}})
\end{align}
with
\begin{align}
\alpha_{\bar{q}} &=\bar{q}\exp\left(-\frac{\theta_{E}}{T}\right),
\end{align}
where $\beta={1}/{k_{B}T}$, $k_B$ is the Boltzmann constant and $\theta_E$ is the
Einstein temperature defined as $\theta_E ={\hbar\omega_E}/{k_B}$.
For later use, it is useful to define a $\bar{q}$-deformed Einstein function $E(\alpha_{\bar{q}})$ as
\begin{eqnarray}\label{eq9.1} E(\alpha_{{\bar{q}}}) = \frac{\alpha_{{\bar{q}}}}{{\bar{q}}}
\left[\frac{\theta_E}{T(\alpha_{{\bar{q}}}-1)}\right]^2,\end{eqnarray}
in the limit ${\bar{q}}\to 1$ we recover the Einstein function $E(\alpha)$ \cite{patt}.
In the following, we shall find several important thermodynamic quantities. First, by using the result of Eq.~(\ref{eq10}) we can determine the  ${\bar{q}}$-deformed entropy 
\begin{eqnarray} s_{{\bar{q}}} = -\frac{\partial f_{{\bar{q}}}}{\partial T} =
-\frac{k_{B}}{{\bar{q}}}\left[\ln(1-\alpha_{{\bar{q}}})-\frac{\theta_E\alpha_{{\bar{q}}}}
{T(1-\alpha_{{\bar{q}}})}\right],\end{eqnarray}

and the ${\bar{q}}$-deformed specific heat can be now determined and it is given by
\begin{eqnarray}c_{V{\bar{q}}}(T) = T\left(\frac{\partial s_{\bar{q}}}{\partial T}\right) =
\frac{k_{B}\alpha_{\bar{q}}}{\bar{q}}\left[\frac{\theta_E}{T(\alpha_{\bar{q}}-1)}\right]^2.\end{eqnarray}
The specific heat of the Einstein solid as a function of $E(\alpha_{\bar{q}})$, defined in Eq.\ (\ref{eq9.1}), can be written as follows
\begin{equation} \label{eq11} c_{V{\bar{q}}}(T) = 3k_{B}E(\alpha_{\bar{q}}).\end{equation}
The complete behaviors are depicted in Figs.\ \ref{grafi2} and \ref{graf2} for $q$-bosons and $B$-anyons, respectively.

In order to make a good comparison between the behavior of deformed functions of both the $q$-boson case and the $B$-anyon case, let us summarize the corresponding results of Ref.\ \cite{bri}. For the $q$-boson case, one should note that when $T\gg\theta_E$, with $\theta_E$ $\sim$ $100 K$ for common crystals, one recovers the classical result
$c_{V{{q}}}\to 3k_B$, for $q=1$, known as the Dulong-Petit law (see Fig.\ \ref{grafi2}(c)). On the other hand, for
sufficiently low temperatures, where $T\ll\theta_E$, the specific heat goes exponentially with temperature \cite{sal} as
\begin{equation}\label{eq15} c_{V{{q}}} \rightarrow k_B
\left(\frac{\theta_E}{T}\right)^2\frac{1}{{{q}}^2}\exp\left(-\frac{\theta_E}{T}\right). \end{equation}
%
However, it is well-known that at sufficiently low temperatures, the specific heat of solids does not experimentally follow the exponential function described in
Eq.\ (\ref{eq15}). Thus, the undeformed Einstein model recovers the usual specific heat of the solids only at high temperatures. Also, we remark that is very clear from Fig.\ \ref{grafi2}(b) that the $q = 0.1$ case (blue dashed line) is the only one that presents a slight change for intermediate temperatures, a situation that is totally opposite to what happens in the present paper. Also from Ref.\ \cite{bri}, the specific heat in a ${q}$-deformed Debye solid behaves as: (i) at low temperature, $T\ll\theta_D$, $c_{V{{q}}}$ is proportional to $T^3$; while for (ii) high temperature, $T\gg\theta_D$, the classical result limit $c_{V{{q}}}\to 3k_B$  is also recovered. Both results are known from the literature.

Now let us consider the $B$-anyons. For the ${\bar{q}}$-deformed case, as shown in Fig.\ \ref{graf2}, we observe that with increasing temperature the black (dot-dashed), green (dashed) and blue (dotted) curves, which represent a greater presence of ${\bar{q}}$-deformation, exhibit significantly different behaviors compared to the red curve (solid) corresponding to ${\bar{q}}$=1. Thus, let us now explore this new picture for
$B$-anyons. At large temperatures the specific heat becomes
\begin{align}
c_{V{\bar{q}}}(T) &\simeq 3k_B \left(\frac{\theta_E}{T}
\right)^2\frac{1}{\left[({\bar{q}}-1)-{\bar{q}}
  \left(\frac{\theta_E}{T}\right)\right]^2}\nonumber \\ &=3k_B
\left(\frac{\theta_E}{T}
\right)^2\frac{1}{\left[\varepsilon+\left(\frac{\theta_E}{T}\right)\right]^2}.\end{align}
In the last step we have rewritten the equation in terms of the
definition of ${\bar{q}}=1-\varepsilon$, with $\varepsilon\ll1$, given in
Subsec.\ \ref{Banyons}. Notice that the specific heat always goes
to zero as $T\gg \theta_E$ for ${\bar{q}}<1$. This is precisely a consequence
of the harmonic oscillator spectrum of Eq.\ (\ref{eq8}) submitted to
the B-anyons statistics of Eq.\ (\ref{e50.1}). Alternatively, one could
also consider the undeformed Bose-Einstein statistics for the spectrum
of Eq.\ (\ref{eq8}) with ${{\bar{q}}\,}^n$ degeneracy for each $n$-th state. Such
degeneracy could be associated with interactions among oscillators in
contrast with the original Einstein consideration of $3N$ independent
harmonic oscillators describing the solid. Thus, the present
intermediate statistics could capture realistic interactions even
though one previously assumes independent oscillators. Another
possibility could be the existence of extra degrees of freedom with
same energy quantum number that could be probed by the $B$-anyons
statistics.

Finally, the most revealing fact is that the specific heat
behaves as the specific heat of a two level system with a maximum at
$T\sim\theta_E$, that is the well-known {\it Schottky effect}, and
since it was previously assumed an infinite number of states, such an
apparent discrepancy clearly can be seen as an effect of a changing of
states counting that may be well understood in terms of the
intermediate statistics. This ``apparent mismatch'' of number of states
can be indeed related to impurities. For instance, it has been
reported long ago \cite{YBACO} that some different samples of the
YBaCuO high-$T_c$ superconductor presented Schottky effect in the
specific heat due to the presence of heavy rare-earth-ion impurities. Similar results were also found in disordered magnetic systems \cite{cbezerra}.

We remark that the $q$- or $\bar{q}$-deformed thermodynamic relationships in the Einstein solid model are obtained from the insertion of the basic
number defined in Eq.\ (\ref{eq2}) into the Hamiltonian given by Eq.\ (\ref{eq1}). Therefore, the results come naturally straightforward and the
results are dependent on this deformation (or disorder) parameter. It is worth to mention a recent work where we associate $q$ with
the nanostructures annealing temperature, modeling the experimental data obtained through this technique of reordering of the
studied alloys \cite{marinho}.

\subsection{Deformed Debye solid}
\label{dds}

Corrections of the Einstein model are given by the Debye model
allowing us to integrate out a continuous spectrum of frequencies up
to the Debye frequency $\omega_D$, giving the total number of normal
vibrational modes \cite{patt,sal,reif,hua,kit},
\begin{equation} \displaystyle\int_{0}^{\omega_D}g(\omega)d\omega = 3N, \end{equation}
where $g(\omega)d\omega$ denotes the number of normal vibrational modes whose frequency is in the range
$(\omega, \omega+d\omega)$. The function $g(\omega)$ can be given in terms of Rayleigh expression as follows
\begin{equation} \label{eq.20}8\pi\left(\frac{1}{\lambda}\right)^{2}d\left(\frac{1}{\lambda}\right) =
\frac{\omega^2 d\omega} {\pi^{2}c^{3}},\end{equation}
where $c$ is the speed of light and $\lambda $ is the wavelength.
The expected energy value of the Planck oscillator with frequency $\omega$ for the $B$-anyons statistics is easily obtained from the
internal energy $U_{\bar{q}}$ excluding the zero point term ${\hbar\omega}/{2}$, i.e.,
\begin{equation} U_{\bar{q}} =
N\left[\frac{\hbar\omega}{\exp\left(\frac{\hbar\omega}{k_{B}T}\right)-{\bar{q}}}
\right]=Nu_{\bar{q}},\end{equation}
with
\begin{equation}\label{eq.21}u_{\bar{q}} = \frac{\hbar\omega}{\exp\left(\frac{\hbar\omega}{k_{B}T}\right)-{\bar{q}}}.
\end{equation}
The explicit steps to obtain Eq.\ (\ref{eq.21}) are given in appendix B.

In order to determine the number of photons between $\omega$ and
$\omega+d\omega$, one makes use of the volume in a region on the phase space \cite{patt},
which results
\begin{eqnarray} \label{eq.22} g(\omega)d\omega \approx \frac{2V}{h^3}\left[4\pi\left(\frac{\hbar\omega}{c}\right)^2
\left(\frac{\hbar d\omega}{c}\right)\right] = \frac{V\omega^{2}d\omega}{\pi^{2}c^{3}}. \end{eqnarray}
We now apply the ${\bar{q}}$-deformation in the same way as in Eq.~(\ref{eq11}), i.e.
\begin{equation} c_{V_{\bar{q}}}(T) = 3k_{B}D(\alpha_{0\bar{q}}), \end{equation}
where $D(\alpha_{0\bar{q}})$ is the ${\bar{q}}$-deformed Debye function that we are going to find shortly. First let us define the function
\begin{equation} \label{eq.23} D(\alpha_{0\bar{q}}) = \frac{3}{\alpha_{0}^3} \int_{0}^{\alpha_{0}}
\frac{\alpha^4\exp{\alpha}}{[\exp(\alpha)-{\bar{q}}]^2}\;d\alpha, \quad
\alpha_{0} = \frac{\hbar\omega_{D}}{k_{B}T} = \frac{\theta_{D}}{T},\end{equation}
with $\omega_{D}$ and $\theta_{D}$ being the Debye characteristic frequency and temperature, respectively.
To perform the integration, it is convenient to make the following change of variable
\be \alpha_{0\bar{q}} = \frac{\alpha_{0}-\ln{\bar{q}}}{{\bar{q}}}. \ee
Consequently, we obtain as a result of the integration
\begin{widetext}
\begin{eqnarray} \label{eq12.5} D(\alpha_{0\bar{q}}) &=& \frac{{\alpha_{0\bar{q}}}}{\rm \exp(\alpha_{0\bar{q}}-\ln{{\bar{q}}})}-1+\frac{12}
{{\bar{q}\,}^4\alpha_{0\bar{q}\,}^3}\Bigg\{\frac{\ln{{\bar{q}\,}^4}}{4}+\ln{{\bar{q}\,}^3}
\ln\Bigg(\frac{\bar{q}-1}{\bar{q}}\Bigg)-\nonumber\\ &-& 3\ln{{\bar{q}\,}^2}\rm{Li_2}\Bigg(\frac{1}{\bar{q}}\Bigg)- 6\ln{{\bar{q}}}\rm{Li_3}\Bigg(\frac{1}{{\bar{q}}}\Bigg)
- 6\rm{Li_4}\Bigg(\frac{1}{\bar{q}}\Bigg)-\frac{{\bar{q}\,}^4}{4}\;\alpha_{0\bar{q}\,}^4+\nonumber\\ &+&
{\bar{q}\,}^3\alpha_{0\bar{q}\,}^3\ln\left[1-\exp(\alpha_{0\bar{q}}-\ln\bar{q})\right]
+ 3{\bar{q}\,}^2\alpha_{0\bar{q}}^2\rm{Li_2}\left[\exp(\alpha_{0\bar{q}}-\ln\bar{q})\right] -\nonumber\\ &-&
 6{\bar{q}}\alpha_{0\bar{q}}\rm{Li_3}\left[\exp(\alpha_{0\bar{q}}-\ln\bar{q})\right]+6\rm{Li_4}\left[\exp(\alpha_{0\bar{q}}-\ln\bar{q})\right]\Bigg\},
\end{eqnarray}
\end{widetext}
where
\be \rm{Li_n(z)}=\displaystyle\sum_{k=1}^{\infty}{\frac{z^k}{k^{n}}} \ee
is the polylogarithm function.

Two important limits must be discussed. First, for $\alpha_{0\bar{q}}\ll 1$, the function $D(\alpha_{0\bar{q}})$ can be
expressed in terms of the deforming parameter $\bar{q}$ of the intermediate statistics, i.e.
\begin{equation} D(\alpha_{0\bar{q}}) =  \frac{\ln\bar{q}}{\bar{q}-1} + \frac{\alpha_{0\bar{q}}}{1-\bar{q}}\;
(13+\ln\bar{q}) + \cdots, \end{equation}
such that at high temperature, $T\gg\theta_D$,  the corresponding specific heat is given by
\begin{eqnarray}\label{DulongPetit}  c_{V_{\bar{q}}}\to 3k_B\;\frac{\ln\bar{q}}{\bar{q}-1}.
\end{eqnarray}
Here we can see that the specific heat behaves asymptotically larger than $3k_B$ for $\bar{q}<1$, which means it is effectively going beyond the Dulong-Petit limit at high temperature. This phenomenon may be related to anharmonicity of interatomic interactions \cite{beyond} and looks to be probed through $B$-anyons statistics. Second, on the other hand, for $\alpha_{0\bar{q}}\gg 1$, we can write the $\bar{q}$-deformed Debye function (\ref{eq12.5}) as
\begin{equation}
D(\alpha_{0\bar{q}})\approx \frac{4\pi^4}{5{[\bar{q}\alpha_{0\bar{q}}]^3}}. \end{equation}
Thus, as in the usual Debye solid, at low temperature, $T\ll\theta_D$, the specific heat in a $\bar{q}$-deformed Debye solid is proportional to $T^3$. This is a consequence of phonon excitation, a fact that is in agreement with experiments. In addition, we express the $\bar{q}$-deformed specific heat at low temperature as
\begin{eqnarray} c_{V_{\bar{q}}} = 3k_{B}D(\alpha_{0\bar{q}})=\frac{12\pi^4k_B}{5[\bar{q}\alpha_{0\bar{q}}]^3}=
\frac{1944}{[\bar{q}\alpha_{0\bar{q}}]^3} \frac{J}{mol K},  \nonumber \\ \end{eqnarray}
which is in complete agreement with the literature for the case without deformation \cite{patt}. We remark that all steps to obtain $D(\alpha_{0\bar{q}})$ were performed above.

Regarding the thermal conductivity $\kappa$, it can be obtained through the usual relationship established with the specific heat \cite{zim}
\begin{equation} \label{eq13}\kappa = \frac{1}{3}c_{V}v l, \end{equation}
where $v$ is the average particle speed and $l$ is the space between particles. We can deduce a relationship between
thermal conductivity $\kappa$ and electrical conductivity $\sigma$ eliminating $l$. We know that $\sigma=\frac{ne^2l}{mv}$,
where $m$, $n$ e $e$, are the mass, number and charge of electrons, respectively. Therefore, we can write
\begin{equation} \frac{\kappa}{\sigma}=\frac{1}{3}\frac{c_{V}mv^2}{ne^2}.\end{equation}
Also, in a classic gas the average energy of a particle is $\frac{1}{2}mv^2=\frac{3}{2}k_{B}T$,
while the specific heat is $\frac{3}{2}nk_{B}$, so
\be\label{eq14} \frac{\kappa}{\sigma}=\frac{3}{2}\left(\frac{k_B}{e}\right)^2T. \ee
The ratio ${\kappa}/{\sigma T}$ is called \textit{Lorenz number} and it is a constant,
independent of temperature and scattering mechanism. This is the Wiedemann-Franz law,
which is satisfied experimentally, and the Lorenz number is then determined \cite{zim}.

In the following we shall explore the $\bar{q}$-deformed (and $q$-deformed) thermal and electrical conductivities which are related with their corresponding undeformed counterparts by \cite{bri}
\begin{eqnarray}\kappa_{\bar{q}}=\kappa \frac{c_{V_{\bar{q}}}}{c_{V}} \qquad\qquad \mbox{and} \qquad\qquad
\sigma_{\bar{q}}=\sigma \frac{\kappa_{\bar{q}}}{\kappa}.\end{eqnarray}

\begin{table}[hbt]
\centering
\begin{tabular}{|c||c|c|c|c||c|c|c|c||c|c|c|c|}
\hline
{Element} &
\multicolumn {4} {c||} { (a) $c_{V_{\bar{q}}}$} &
\multicolumn {4} {c||} { (b) $\kappa_{\bar{q}}$} &
\multicolumn {4} {c|} { (c) $\sigma_{\bar{q}}$} \\
\cline{2-13}
& {$\overline{q}$=1.0} & {$\overline{q}$=0.9} & {$\overline{q}$=0.5} &
{$\overline{q}$=0.1} & {$\overline{q}$=1.0} & {$\overline{q}$=0.9} & {$\overline{q}$=0.5}
& {$\overline{q}$=0.1} & {$\overline{q}$=1.0} & {$\overline{q}$=0.9} & {$\overline{q}$=0.5} & {$\overline{q}$=0.1} \\
\hline
Rb & 3.0$\times10^5$ & 7.8$\times10^4$ & 2.9$\times10^3$ & 126 & 0.58 & 0.15 & 0.006 & 2.4$\times10^{-4}$
& 8.0$\times10^4$ & 2.1$\times10^4$ & 764 & 34 \\
\hline
Pb & 4.5$\times10^4$ & 2.1$\times10^4$ & 1.7$\times10^3$ & 104 & 0.35 & 0.16 & 0.013 & 8.0$\times10^{-4}$
& 4.8$\times10^4$ & 2.2$\times10^4$ & 1.8$\times10^3$ & 110 \\
\hline
Bi & 3.1$\times10^4$ & 1.5$\times10^4$ & 1.5$\times10^3$ & 99 & 0.08 & 0.04 & 0.039 & 2.5$\times10^{-4}$
& 8.6$\times10^3$ & 4.2$\times10^3$ & 415 & 27 \\
\hline
Sn & 6.6$\times10^3$ & 4.2$\times10^3$ & 773 & 44 & 0.67 & 0.43 & 0.078 & 0.008 & 9.1$\times10^4$
& 5.9$\times10^4$ & 1.1$\times10^4$ & 1.0$\times10^3$ \\
\hline
Pt & 3.8$\times10^3$ & 2.6$\times10^3$ & 584 & 65 & 0.72 & 0.50 & 0.114 & 0.012 & 9.6$\times10^4$
& 6.6$\times10^4$ & 1.5$\times10^4$ & 1.6$\times10^3$ \\
\hline
Ga & 1.6$\times10^3$ & 1.2$\times10^3$ & 357 & 51 & 0.41 & 0.31 & 0.092 & 0.013 & 6.7$\times10^4$
& 5.1$\times10^4$ & 1.5$\times10^4$ & 1.6$\times10^3$ \\
\hline
Li & 1.3$\times10^3$ & 991 & 312 & 47 & 0.85 & 0.65 & 0.203 & 0.032 & 1.1$\times10^5$
& 8.5$\times10^4$ & 2.6$\times10^4$ & 4.0$\times10^3$ \\
\hline
Ti & 708 & 570 & 228 & 41 & 0.46 & 0.37 & 0.142 & 0.025 & 2.3$\times10^4$ & 1.8$\times10^4$ & 6.9$\times10^3$
& 1.2$\times10^3$ \\
\hline
Fe & 506 & 416 & 168 & 34 & 0.80 & 0.66 & 0.268 & 0.053 & 1.0$\times10^5$ & 8.2$\times10^4$ & 3.4$\times10^4$
& 6.8$\times10^3$ \\
\hline
Rh & 475 & 392 & 161 & 33 & 1.5 & 1.3 & 0.513 & 0.102 & 2.1$\times10^5$ & $1.7\times10^5$ & 7.1$\times10^4$ & $1.4\times10^4$ \\
\hline
Os & 420 & 349 & 148 & 31 & 0.88 & 0.732 & 0.314 & 0.072 & 1.1$\times10^5$ & 9.1$\times10^4$ & 3.9$\times10^4$
& 8.1$\times10^3$ \\
\hline
Ru & 243 & 208 & 100 & 24 & 1.17 & 1.08 & 0.478 & 0.123 & 1.4$\times10^5$ & 1.2$\times10^5$ & 5.5$\times10^4$ & 1.4$\times10^4$ \\
\hline
Cr & 210 & 181 & 89 & 23 & 0.94 & 0.81 & 0.404 & 0.102 & 7.8$\times10^4$ & 6.7$\times10^4$ & 3.3$\times10^4$ & 8.5$\times10^3$ \\
\hline
Si & 196 & 169 & 85 & 22 & 1.48 & 1.28 & 0.643 & 0.174 & - & - & - & - \\
\hline
\end{tabular}
\caption{Chemical elements and their respective $\bar{q}$-deformed: (a) specific heat $c_{V_{\bar{q}}}$ {$\left(\frac{J}{mol K}\right)$},
(b) thermal conductivity {$\kappa_{\bar{q}}$} {$\left(\frac{W}{cm\dot K}\right)$},
and (c) electrical conductivity {$\sigma_{\bar{q}}$} {$(ohm\cdot cm)$}, for $T = 300K$ and $\bar{q}=1.0$, $0.9$, $0.5$ and $0.1$
\cite{kit}. Discussion is in the main text.}
\label{tab2}
\end{table}

\noindent In Tab.\ \ref{tab2} we present the specific heat, thermal and electrical conductivity for some chemical elements, after application of the deformation. We have (i) the undeformed case ($\bar{q}=1.0$); (ii) a slight deformation case ($\bar{q} = 0.9$); (iii) an intermediate deformation case ($\bar{q} = 0.5$); and  (iv) a strong deformation case ($\bar{q} = 0.1$). For illustration purposes, we choose iron (Fe) and chromium (Cr), which are two materials that can be employed in many scientific areas of interest.

\begin{figure}[htb!]
\centerline{
\includegraphics[{angle=90,height=8.0cm,angle=270,width=10.0cm}]{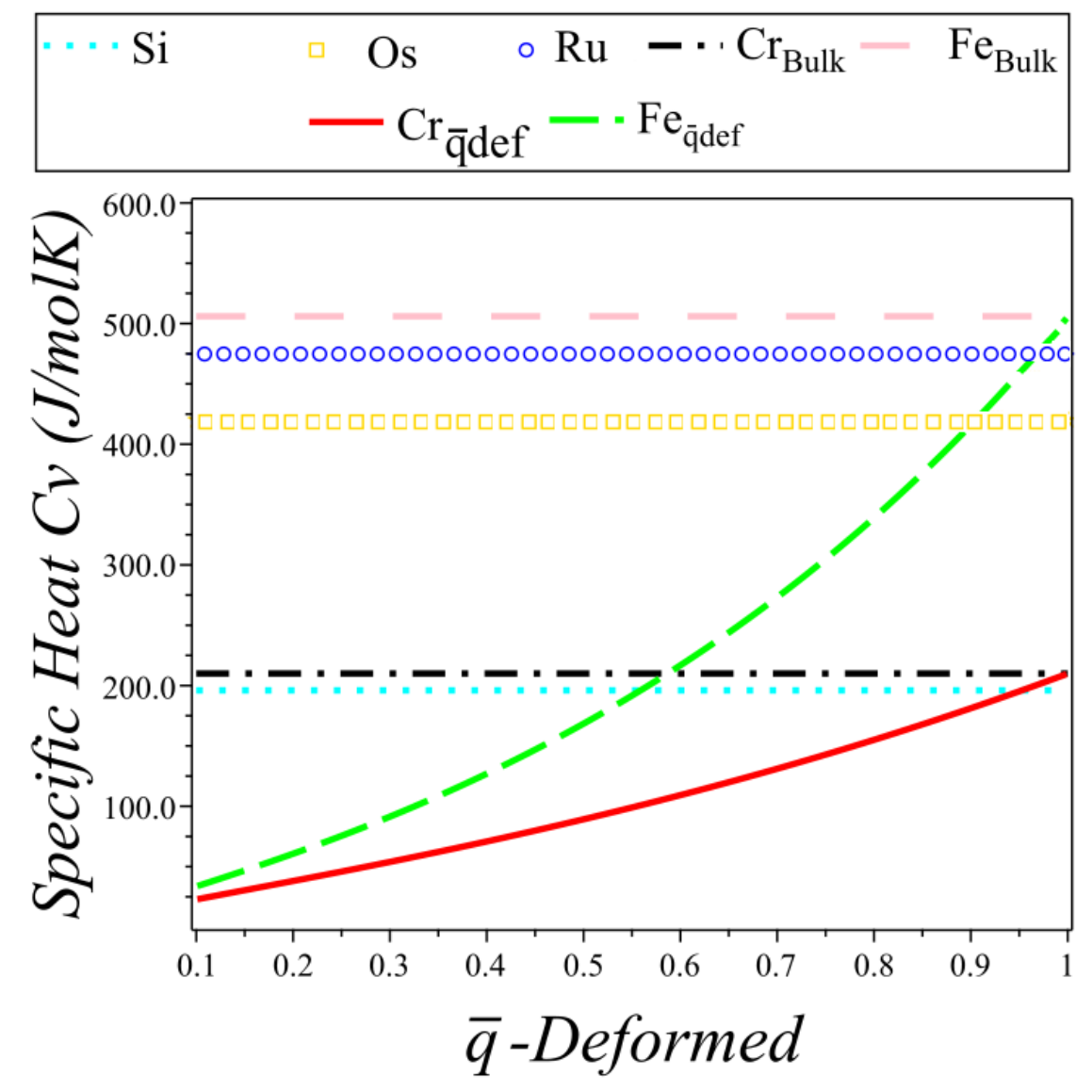}
}\caption{$\bar{q}$-deformed specific heat $c_{V_{\bar{q}}}$, for Fe and Cr as a function of $\bar{q}$, from $\bar{q}=0.1$ to $\bar{q} = 1$. Discussion is in the main text.}
\label{graf3}
\end{figure}

\begin{figure}[htb!]
\centerline{
\includegraphics[{angle=90,height=8.0cm,angle=270,width=10.0cm}]{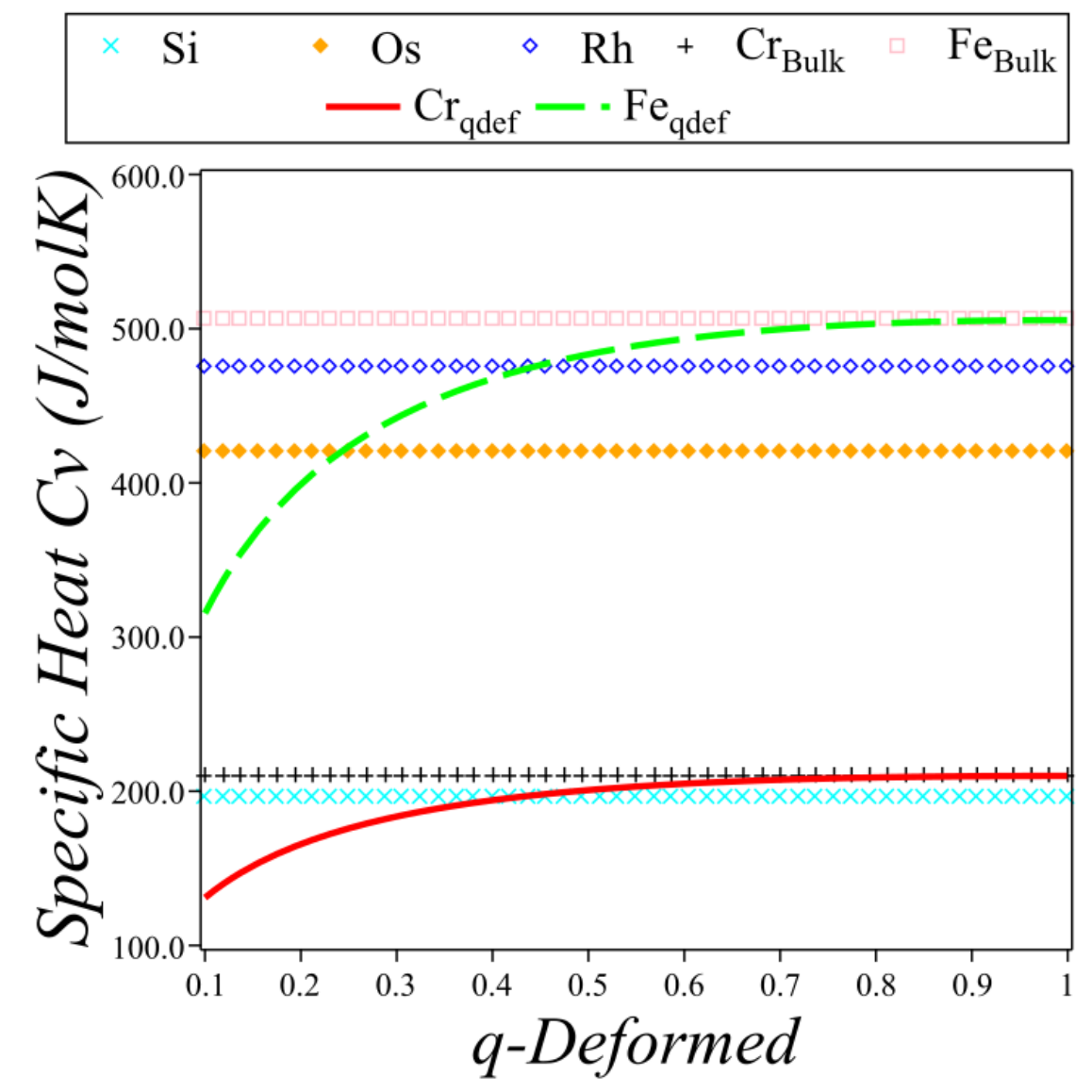}
}\caption{$q$-deformed specific heat $c_{V_q}$, for Fe and Cr as a function of $q$, from $q=0.1$ to ${q} = 1$. See Ref.\ \cite{bri}. Discussion is in the main text.}
\label{grafi3}
\end{figure}

\begin{figure*}[htb!]
\centerline{
\includegraphics[{angle=90,height=8.0cm,angle=270,width=9.0cm}]{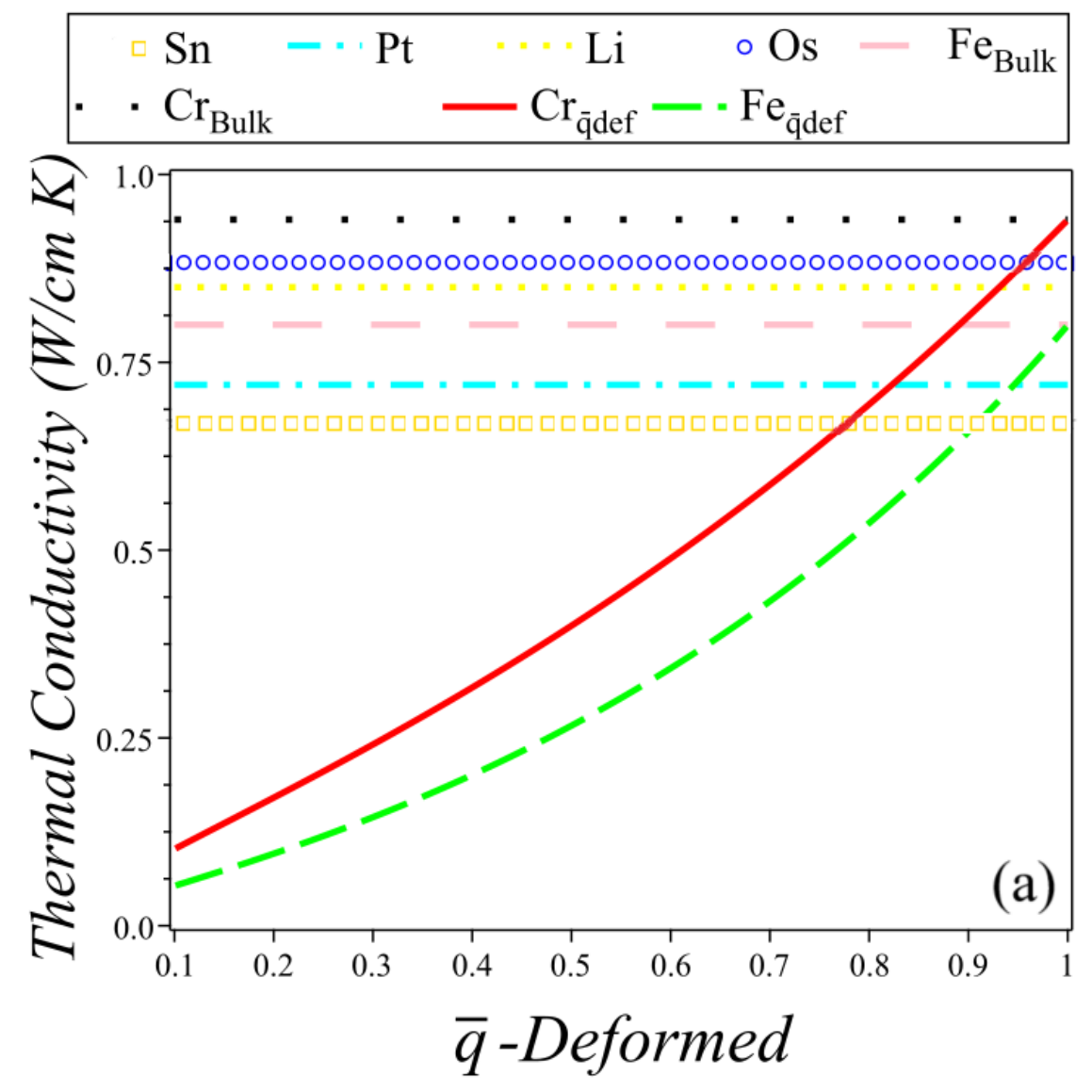}
\includegraphics[{angle=90,height=8.0cm,angle=270,width=9.0cm}]{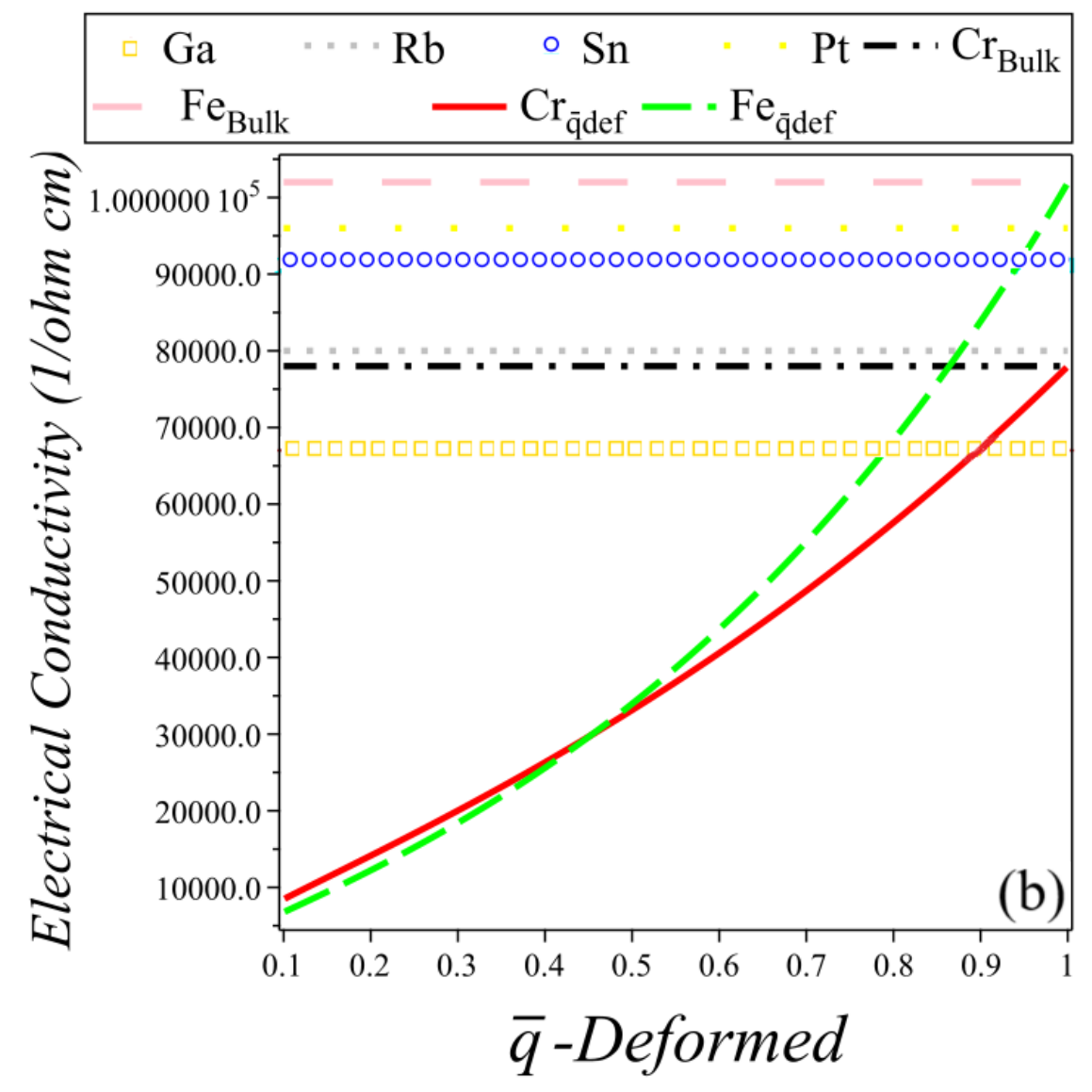}
}\caption{(a) $\bar{q}$-deformed  thermal conductivity $\kappa_{\bar{q}}$ and (b) $\bar{q}$-deformed electrical conductivity $\sigma_{\bar{q}}$, for Fe and Cr as functions of $\bar{q}$, from $\bar{q}=0.1$ to $\bar{q} = 1$. Discussion is in the main text.}
\label{graf5}
\end{figure*}

\begin{figure*}[htb!]
\centerline{
\includegraphics[{angle=90,height=8.0cm,angle=270,width=9.0cm}]{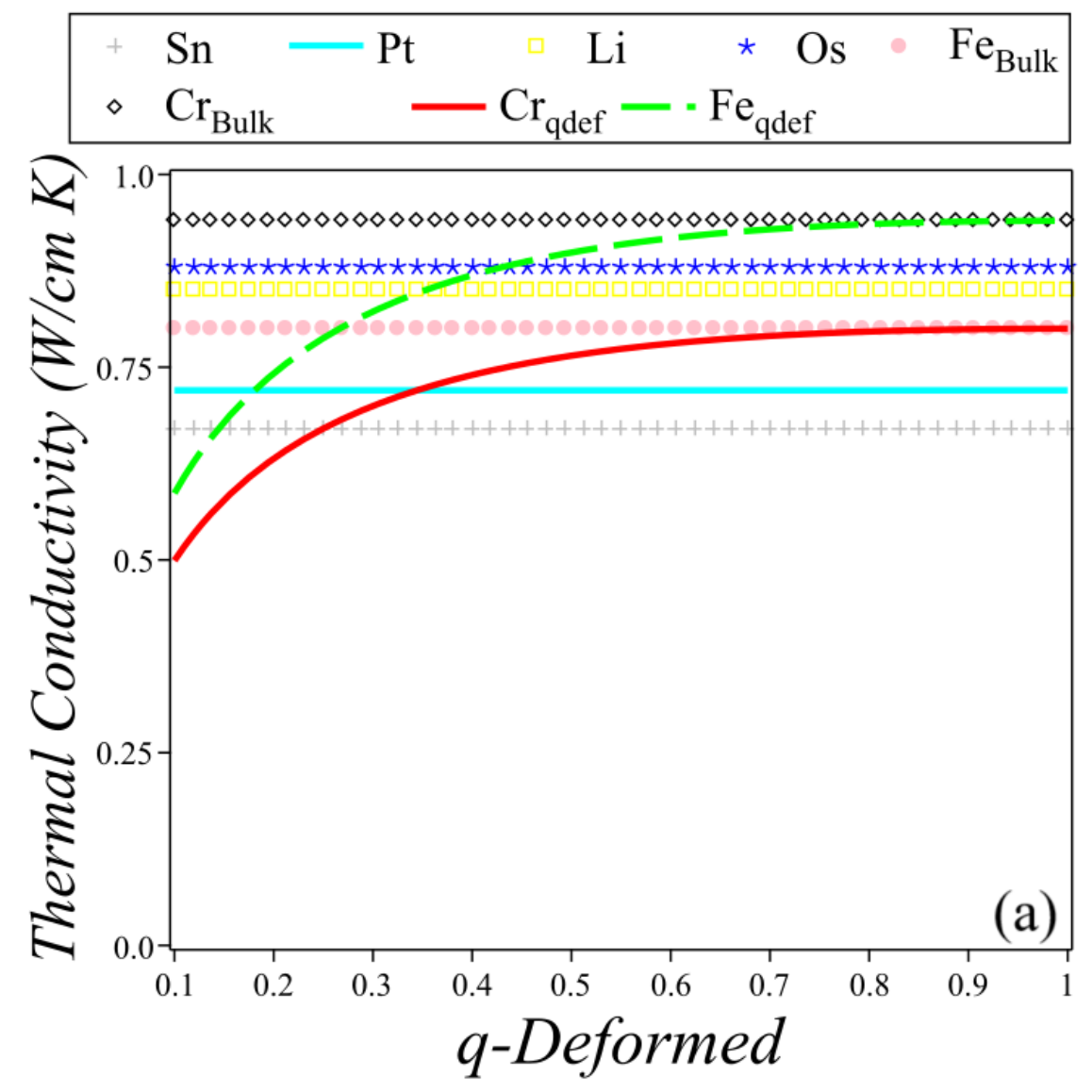}
\includegraphics[{angle=90,height=8.0cm,angle=270,width=9.0cm}]{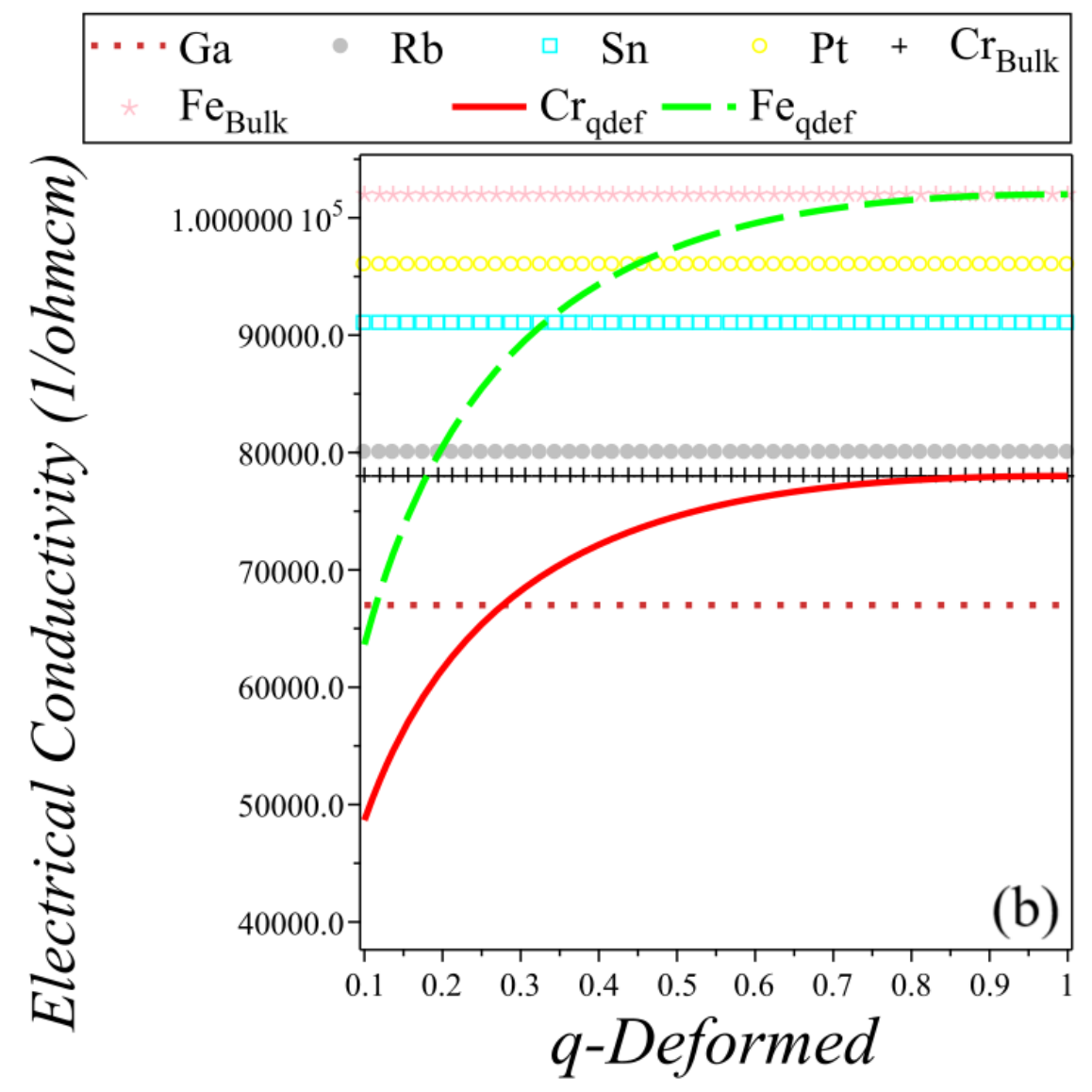}
}\caption{(a) ${q}$-deformed  thermal conductivity $\kappa_{{q}}$ and (b) ${q}$-deformed electrical conductivity $\sigma_{{q}}$, for Fe and Cr as functions of ${q}$, from ${q}=0.1$ to ${q} = 1$. See Ref.\ \cite{bri}. Discussion is in the main text.}
\label{grafi5}
\end{figure*}

\begin{figure}[htb!]
\centerline{
\includegraphics[{angle=90,height=8.0cm,angle=270,width=10.0cm}]{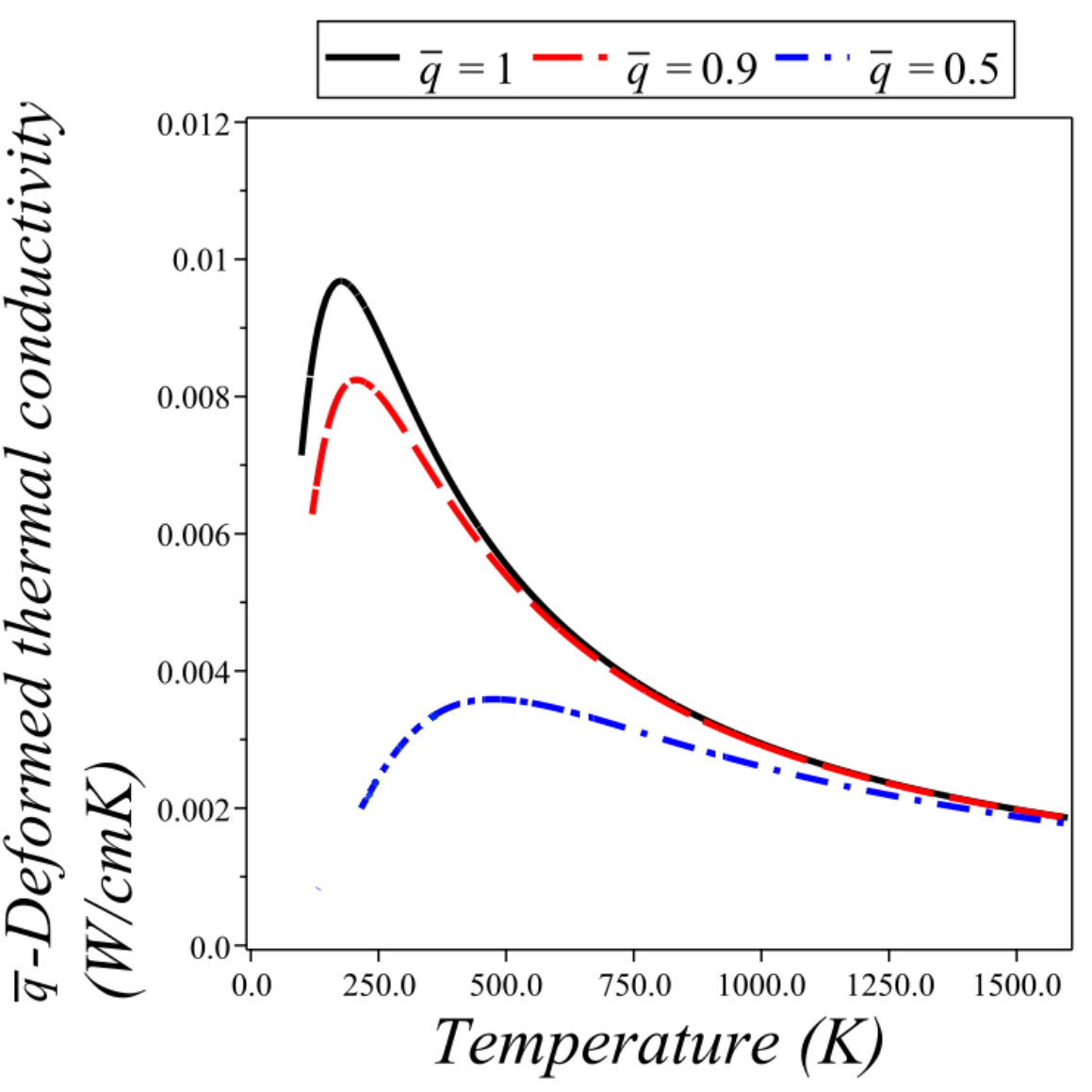}
}\caption{$\bar{q}$-deformed thermal conductivity $\kappa_{\bar{q}}$, for $\rm Cr_{Bulk}$
($\bar{q}=1$) and $\rm Cr_{\bar{q}def}$ ($\bar{q}=0.5$ and $0.9$), as a function of temperature from $T = 0K$ to $T = 1400K$. Discussion is in the main text.}
\label{graf6}
\end{figure}

\begin{figure}[htb!]
\centerline{
\includegraphics[{angle=90,height=8.0cm,angle=270,width=10.0cm}]{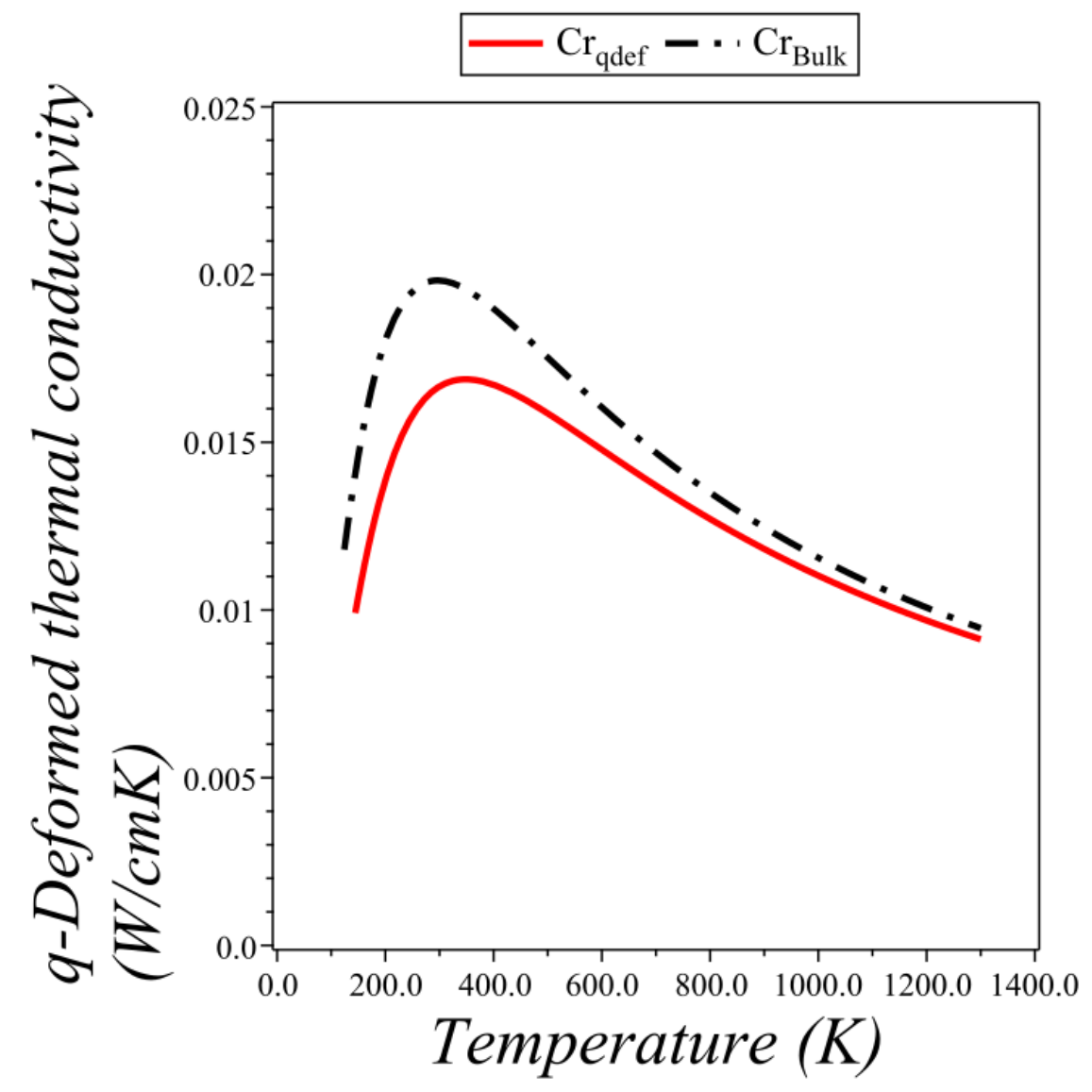}
}\caption{${q}$-deformed thermal conductivity $\kappa_{{q}}$, for $\rm Cr_{Bulk}$
(${q}=1$) and $\rm Cr_{{q}def}$ (${q}=0.1$), as a function of temperature from $T = 0K$ to $T = 1400K$. See Ref.\ \cite{bri}. Discussion is in the main text.}
\label{grafi6}
\end{figure}

We can observe the behavior of the specific heat of iron (Fe) (green dashed curve in Fig.\ \ref{graf3}), as a function of the deformation parameter, starting from the case without deformation ($Fe_{Bulk}$, i.e., when $\bar{q}=1$).  As the parameter $\bar{q}$ is acting we notice that this curve reaches the specific heat of other elements, such as the osmium (Os) when $\bar{q}\approx 0.9$ and the $Cr_{Bulk}$ at $\bar{q}\approx 0.58$. For the sake of comparison, in Fig.\ \ref{grafi3}, is shown the behavior for $q$-bosons, which asymptotically is completely different from the $B$-anyons case. Whereas in the former case the curves tend to overlap for large $q$ (small deformation), in the latter case the curves tend to overlap for small $\bar{q}$ (large deformation). As expected, since they depend linearly with the specific heat, the deformed thermal and electrical conductivities follow similar behavior, as shown in Figs.\ \ref{graf5} and \ref{grafi5}, i.e., the curves degenerate at opposite deformation limits depending on the case involved: $q$-bosons or $B$-anyons. We also emphasize the overlap clearly shown in Fig.\ \ref{graf5}(b) of the deformed electrical conductivity curves that starts at
$\bar{q}\approx 0.47$. 
On the other hand, we plot in Figs.\ \ref{graf6} and \ref{grafi6} the behavior of the thermal conductivity of deformed chromium (solid red line)
(as in enriched or doped samples \cite{kit}) as a function of temperature. It is possible to verify that as
the temperature increases the behavior of the curves becomes similar, but they overlap more slowly for $B$-anyons, Fig.\ \ref{graf6}, than for $q$-bosons, Fig.\ \ref{grafi6}, as expected according to our previous discussions.



\section{Conclusions}
\label{con}

In summary, in this work we study the $\bar{q}$($q$)-deformed Einstein and Debye models and the associated $\bar{q}$($q$)-deformed specific heat, as well as the $\bar{q}$($q$)-deformed thermal and electrical conductivities, by using the mathematics of $q$-analogs. We revisited the results for $q$-bosons \cite{bri} and obtained the results for $B$-anyons. On one hand, for the $q$-bosons case our results are as follows: (i) for the Einstein model, when $T\gg\theta_E$, the Dulong-Petit law $c_{V{{q}}}\to 3k_B$ is recovered. However, for low temperatures, where $T\ll\theta_E$, the specific heat goes exponentially with temperature. (ii) For the Debye model, the specific heat behaves as: at low temperature, $T\ll\theta_D$, $c_{V{{q}}}$ is proportional to $T^3$; while for high temperature, $T\gg\theta_D$, the classical limit $c_{V{{q}}}\to 3k_B$ is also recovered. We remark that these results are in agreement with Marinho and collaborators \cite{bri}. On the other hand, for the $B$-anyons case the scenario is a little different: (i) for the Einstein model, when $T\gg\theta_E$, the specific heat always goes to zero. More interesting, however, the specific heat behaves as the specific heat of a two level system with a maximum at $T\sim\theta_E$, that is the well-known {\it Schottky effect}, which can be related to impurities or disorder \cite{YBACO,cbezerra}. (ii) For the Debye model, at low temperature, $T\ll\theta_D$, the specific heat in a $\bar{q}$-deformed Debye solid is proportional to $T^3$. This is a consequence of phonon excitation. At high temperature, $T\gg\theta_D$, the specific heat behaves asymptotically larger than $3k_B$, which means it is effectively going beyond the Dulong-Petit limit at high temperature.

The numerical results obtained in the present work allow us several important observations. One of them is the fact that in the $q$-deformation case there is a region of larger values of the deforming parameter $q$ (less deformed regime) in which different curves of thermal and electrical conductivities coincide, that is, the curves degenerate much faster than in the case of $B$-anyons and vice-versa. On the other side, i.e., in the region of small values of $\bar{q}$, the curves of thermal and electrical conductivities degenerate approaching zero. This leads to {\it completely different behaviors}. Alternatively, since in the $q$-bosons the statistics are in principle maintained, the effect of the deformation acts more slowly due to a small change in the crystal lattice. Conversely, B-anyons are part of modified statistics and exhibit greater sensitivity to $\bar{q}$-deformation. Of course, the case of study will decide for one or another deformation type, according to several corresponding situations. For instance, in the $B$-anyons case, smaller values of $\bar{q}$, i.e., large deformations, implies a larger asymptotic value of the specific heat as stated in Eq.\ (\ref{DulongPetit}), which involves physics beyond the Dulong-Petit limit, a phenomenon that may be related to anharmonicity of interatomic interactions \cite{beyond} and looks to be probed through $B$-anyons statistics.

In general, impurities and disorder have the effect of breaking translational invariance symmetry, what may lead to loss of propagation of certain modes and hence to localization effects. The overall effect of these impurities affects the thermal conductivity and other properties in a particular way. Now, the fascinating observation is that $q$($\bar{q}$)-deformation leads to effects that are similar in the specific heat, as well as in the thermal and electrical conductivities, as seen in this present manuscript. Although, the mechanisms are quite different. The deformation only acts through a change in the mean occupation numbers. Still, somehow, the macroscopic effect is similar to that of impurities. We point out two {\it take home messages}: (i) first, we have shown that the deformation acts in a manner somewhat similar to disorder or impurity, modifying the characteristics of a crystalline structure, which are phenomena described by $q$-bosons, and, furthermore, the deformation can also manifest as intermediate statistics, the $B$-anyons (or $B$-type systems); (ii) second, the algebra of $q$-analogs has been applied to the Einstein and Debye models in a mathematically consistent way. No doubt, it should be possible to obtain analogous results for many other important physical models.

In a broader perspective, our results show that scenarios with $\bar{q}$ (or $q$)-deformation develop an interesting mechanism that can be associated with factors that modify certain characteristics of the chemical elements. These interesting scenarios may be associated with the doping or enrichment of the material through the introduction of specific impurities. Moreover, It is worth to mention a recent work where we associate $q$-deformation with the nanostructures annealing temperature \cite{marinho}, modeling the experimental data obtained through this technique of reordering of the studied alloys. Ultimately, this study offers an promising (if incipient) theoretical framework, allowing us to calculate and analyze various thermodynamic properties of the materials before employing any experimental techniques. This $q$-analog based approach paves the way for more informed and precise theoretical investigations that may perhaps also motivate new experimental studies. Surely our model can be realized experimentally and we hope that experimentalists are encouraged to investigate it.

\section*{Note Added}

Recently, the work of Chung and Algin \cite{lastone} was brought to our attention. In that paper, the authors applied the Tamm-Dancoff type $q$-deformed boson algebra to study, among other applications, the Debye solid. In short, they have found that: (i) in high temperatures the specific heat also presents a behavior which differs from the usual Dulong-Petit theory, inclusive with unusual negative values as a function of $q$, associated by the authors with studies on understanding the evolution of stars for the analysis of astrophysical objects;  and (ii) in low temperatures the specific heat presents the usual Debye solid behavior, i.e., it is directly proportional to a term containing $T^3$. However, the values of the $q$-deformed specific heat are higher than the results for the undeformed specific heat.


\section*{Acknowledgments}

AAM acknowledges support from PNPD/CAPES. GMV acknowledges support from CNPq (Grant no.\ 302414/2022-3). FAB acknowledges support from CNPq (Grant no.\ 309092/2022-1). CGB acknowledges support from CNPq (Grant no.\ 309495/2021-0). We also would like to thank the two anonymous referees for their valuable comments/remarks which helped to improve the manuscript.

\appendix

\section{q-deformed mean occupation number}

\noindent The mean value of a physical observable $A$ is given by
\begin{equation}
<A > = tr(\rho A),
\end{equation}
where $\rho$ is the density operator
\begin{equation}
\rho = \frac{\exp(-\beta {\cal H})}{\Xi}.
\end{equation}
Here, $\Xi$ is the grand canonical partition function defined as
\begin{equation}
\Xi = tr\left[\exp(-\beta {\cal H})\right],
\end{equation}
with $\beta=\frac{1}{k_B T}$, where $k_B$  is the Boltzmann constant. The mean occupation number is then calculated by

\begin{equation}
[n_{i}]\label{e34}\equiv \langle[{N_i}]\rangle= {tr}\left[\rho b_{i}^{\dagger}b_i\right] = \frac{{tr}\left[\exp(-\beta {\cal H})
b_{i}^\dagger b_{i}\right]}{\Xi},
\end{equation}

i.e.,
\begin{equation}
[n_{i}] = \frac{{tr}\left\{\exp\left[-\beta\displaystyle\sum_{n_j}
{(E_j-\mu)n_j b_{i}^\dagger b_{i}}\right]\right\}}{{tr}\left\{\exp\left[-\beta\displaystyle\sum_{n_j}{(E_j-\mu)n_j }
\right]\right\}}.
\end{equation}
Applying the cyclic property of the trace
\ben \label{e36}& &tr\left\{\exp\left[-\beta\displaystyle\sum_{n_j}{(E_j-\mu)n_j b_{i}^\dagger b_{i}}\right]\right\}=\nonumber\\
&=&tr\left\{b_{i}^\dagger \exp\left[-\beta\displaystyle\sum_{n_j}{(E_j-\mu)(n_j+\delta_{ji}) b_{i}}\right]\right\}=\nonumber\\
&=&\exp\left[-\beta(E_i-\mu)\right] tr\left\{\exp\left[-\beta\displaystyle\sum_{n_j}{(E_j-\mu)
n_j b_{i}b_{i}^\dagger}\right]\right\},\een
we can write
\ben [n_{i}] = \frac{\exp\left[-\beta(E_i-\mu)\right]tr\left\{\exp\left[-\beta\displaystyle\sum_{n_j}
{(E_j-\mu)n_j b_{i}b_{i}^\dagger }\right]\right\}}{tr\left\{\exp\left[-\beta\displaystyle\sum_{n_j}{(E_j-\mu)n_j }
\right]\right\}}.\een
Using the definition $b_i b_{i}^\dagger=[1+n_i]$, we get
\ben \label{e37}[n_i] = \exp(-\beta(E_i-\mu))[1+n_i]\,\, \rightarrow \,\, \frac{[n_i]}
{[1+n_i]} = z\exp(-\beta E_i).\een
Here $z=\exp(\beta\mu)$ is the fugacity. Solving Eq.(\ref{e37})

\ben [n_{i}] = \frac{z\exp(-\beta E_i)q^{-n_i}}{1-qz\exp(-\beta E_i)} = \frac{q^{-n_i}}
{z^{-1}\exp(\beta E_i)-q},\een

and using the definition of basic number [Eq.\ (4)], we have
\ben \frac{q^{n_i}-q^{-n_i}}{q-q^{-1}} = \frac{q^{-n_i}}{z^{-1}\exp(\beta E_i)-q},\een
or
\ben q^{2n_i} = \frac{z^{-1}\exp(\beta E_i)-q^{-1}}{z^{-1}\exp(\beta E_i)-q}.\een
Finally, we obtain the $q$-deformed mean occupation number,
\ben n_i^{q} = \frac{1}{2\ln q}\ln\left[\frac{z^{-1}\exp(\beta E_i)-q^{-1}}{z^{-1}\exp(\beta E_i)-q}\right].\een

\section{The expected energy value of the Planck oscillator}

From thermodynamics we know the relation between the Helmholtz free energy, entropy and internal energy,
\begin{equation}
f_{\bar{q}}=u_{\bar{q}}-Ts_{\bar{q}}
\end{equation}
or
\begin{equation}
u_{\bar{q}}=f_{\bar{q}}+Ts_{\bar{q}}.
\end{equation}
We know that the Helmholtz free energy is
\be
f_{\bar{q}}=\frac{k_{B} T}{\bar{q}}\ln(1-\alpha_{\bar{q}}),
\ee
and the entropy is
\be
 s_{{\bar{q}}} =-\frac{k_{B}}{{\bar{q}}}\left[\ln(1-\alpha_{{\bar{q}}})-\frac{\theta_E\alpha_{{\bar{q}}}}{T(1-\alpha_{{\bar{q}}})}\right].
\ee
Thus, the internal energy is given by
\begin{align}
  u_{\bar{q}}&=
  f_{\bar{q}} + Ts_{\bar{q}}
  \\
  &
  =
  \frac{k_{B} T}{\bar{q}}\ln(1-\alpha_{\bar{q}})
  +
  T
  \left\{
-\frac{k_{B}}{{\bar{q}}}\left[\ln(1-\alpha_{{\bar{q}}})-\frac{\theta_E\alpha_{{\bar{q}}}}{T(1-\alpha_{{\bar{q}}})}\right]
    \right\}
  \\
  &
  =
{k_B T\over {\bar{q}}}
{\theta_E \alpha_{\bar{q}} \over {T(1-\alpha_{{\bar{q}}}) }}.
\end{align}
By using $\theta_E = {\hbar \omega_E\over k_B}$, we get
\be
u_{\bar{q}}=
{k_B T\over {\bar{q}}}
{ \alpha_{\bar{q}} \hbar \omega_E / k_B  \over {T(1-\alpha_{\bar{q}}) }},
\ee
which may be simplified to
\be
u_{\bar{q}}=
{ \alpha_{\bar{q}} \hbar \omega_E    \over {\bar{q}} {(1-\alpha_{\bar{q}}) }}.
\ee
Now dividing by $\alpha_{\bar{q}}$ we have
\be
u_{\bar{q}}=
{ \hbar \omega_E    \over {\bar{q}}/\alpha_{\bar{q}}}.
\ee
Finally, we obtain
\be
u_{\bar{q}}=
  \left[
  \frac{\hbar\omega}{\exp\left(\frac{\hbar\omega}{k_{B}T}\right)-{\bar{q}}}
\right].
\ee


\end{document}